\documentclass{article}

\usepackage[english]{babel}
\usepackage{indentfirst}

\usepackage[letterpaper,top=2cm,bottom=2cm,left=3cm,right=3cm,marginparwidth=1.75cm]{geometry}

\usepackage{amsmath}
\usepackage{graphicx}
\usepackage[colorlinks=true, allcolors=blue]{hyperref}

\usepackage{authblk}
\title{\textbf{Real-time Manipulation of Liquid Droplets using
Photo-responsive Surfactant}}
\author[a,1]{Xichen Liang}
\author[b,1]{Kseniia M. Karnaukh} 
\author[c]{Lei Zhao}
\author[b]{Serena Seshadri}
\author[b]{Austin J. DuBose}
\author[b]{Sophia J. Bailey}
\author[d]{Qixuan Cao}
\author[c]{Marielle Cooper}
\author[c]{Hao Xu}
\author[a]{Michael Haggmark}
\author[a]{Matthew E. Helgeson}
\author[a]{Michael Gordon}
\author[c,2]{Paolo Luzzatto-Fegiz}
\author[b,2]{Javier Read de Alaniz}
\author[c,2]{Yangying Zhu}

\affil[a]{Department of Chemical Engineering, University of California at Santa Barbara, Santa Barbara, California 93106-5070, USA}
\affil[b]{Department of Chemistry, University of California at Santa Barbara, Santa Barbara, California 93106-5070, USA}
\affil[c]{Department of Mechanical Engineering, University of California at Santa Barbara, Santa Barbara, California 93106-5070, USA}
\affil[d]{Department of Physics, University of California at Santa Barbara, Santa Barbara, California 93106-5070, USA}
\affil[1]{X.L.and K.K. contributed equally to this work.}
\affil[2]{To whom correspondence should be addressed. E-mail: pfegiz@ucsb.edu, jalaniz@ucsb.edu, yangying@ucsb.edu}

\begin{document}
\maketitle

\begin{abstract}
Fast and programmable transport of liquid droplets on a solid substrate is desirable in microfluidic, thermal, biomedical, and energy devices. 
Past research has focused on designing substrates with asymmetric structures or gradient wettability where droplet behaviors are passively controlled, or by applying external electric, thermal, magnetic, or acoustic stimuli that either require the fabrication of electrodes or a strong applied field. 
In this work, we demonstrate tunable and programmable droplet motion on liquid-infused surfaces (LIS) and inside solid-surface capillary channels using low-intensity light and photo-responsive surfactants. 
When illuminated by the light of appropriate wavelengths, the surfactants can reversibly change their molecular conformation thereby tuning interfacial tensions in a multi-phase fluid system. 
This generates a Marangoni flow that drives droplet motions. With two novel surfactants that we synthesized, we demonstrate fast linear and complex 2D movements of droplets on liquid surfaces, on LIS, and inside microchannels.
We also visualized the internal flow pattern using tracer particles and developed simple scaling arguments to explain droplet-size-dependent velocity. 
The method demonstrated in this study serves as a simple and exciting new approach for the dynamic manipulation of droplets for microfluidic, thermal, and water harvesting devices.
\end{abstract}

\section*{Introduction}

Manipulating droplets on surfaces and interfaces is critical for the efficiency and tunability of water desalination \cite{wang2021high}, condensation \cite{cha2020dropwise,hoque2022life,park2016condensation,miljkovic2013jumping}, liquid transport in microfluidics \cite{li2019ionic}, and digital bioassay \cite{honda2021multidimensional}. For example, condensation, a key process in power generation and thermal management, requires fast removal of droplets from the substrates for efficient heat transfer \cite{park2016condensation,oh2020enhanced}. Passive methods to transport droplets on surfaces have focused on engineering asymmetric surface structures \cite{chen2016continuous,chen2018ultrafast,feng2020tip,zhuang2021architecture,launay2020self,feng2021three,lv2014substrate} and gradient chemistry \cite{chaudhury1992make,bai2014efficient,daniel2001fast,deng2017self}. However, these surfaces typically require complex fabrication and do not offer real-time control capability. Active methods that can dynamically tune droplet motions have been developed using electric \cite{li2019ionic,cho2015turning,annapragada2011dynamics,liu2021external}, magnetic \cite{zhu2014real,lei2018high,li2020programmable, khalil2014active}, acoustic \cite{doi:10.1126/sciadv.adg2352}, chemical \cite{zarzar2011bio,he2012synthetic}, thermal \cite{yoshida1995comb,bjelobrk2016thermocapillary}, and photo-thermal \cite{yang2018systems} stimuli. While electric, magnetic and photo-thermal systems may require a high voltage (100-1000 V for electrowetting), magnetic field ($\sim$ 0.1 T), or light intensity ($\sim$ 1 W/mm\textsuperscript{2}) to actuate droplet motion, chemical and thermal systems can have a slow response due to mass diffusion and thermal mass. To realize spatial control, some of these methods still require patterning of microstructures, which limits their spatial resolution \cite{liang2022manipulation}.

Manipulation methods based purely on light stimuli are attractive as light offers spatial resolution down to the diffraction limit, can be easily reconfigured, and does not require the fabrication of microstructures such as electrode arrays \cite{baigl2012photo}. Several studies have investigated photo-sensitive substrates such as TiO\textsubscript{2} or ZnO \cite{wang1997light,sun2001photoinduced}. Although the contact angle of water on these surfaces can decrease by 70-100$^{\circ}$ upon UV illumination, direct manipulation of droplet movement has not been reported.  Optical tweezers based on the transfer of photon momentum have been extensively studied to trap and manipulate solid microparticles \cite{hoffman2015encyclopedia}. They have also not been used to control liquid droplets directly.  Alternatively, photo-responsive surfactants have recently been applied to manipulate multi-phase fluid systems. These are surfactants that can switch their molecular conformation when illuminated by light with appropriate wavelengths. The photoswitch of surfactants results in a local change of the surface tension or interfacial tension, which can further generate a Marangoni flow. Compared to photo-thermal effects (i.e., using light to heat fluid for liquid motion based on Marangoni-flow or temperature-dependent surface tension force), photo-responsive surfactants require a significantly lower light intensity to generate a comparable surface tension change \cite{zhao2022depinning}, and no light-absorbing substrates or nanoparticle additives are needed. Using photo-responsive surfactants, recent works have demonstrated dynamically reconfigurable emulsions \cite{zarzar2015dynamically}, trapping and manipulation of solid objects \cite{lv2018controlling}, moving droplets immersed in \cite{xiao2018moving} and on the surface of an immiscible liquid \cite{xiao2018moving,diguet2009photomanipulation}, and bubble departure from solid surfaces \cite{zhao2022depinning}. However, moving liquid droplets on solid or liquid-infused surfaces using photo-responsive surfactants has not been achieved so far, which is of particular practical interest to many applications.

In this work, we demonstrate fast and programmable 2D motions of liquid droplets on liquid-infused surfaces (LIS), inside solid-wall microchannels, and on liquid substrates using two types of light-sensitive surfactants (SP-DA-PEG and MCH-para) that can induce large changes in the interfacial tension in the opposite directions. The surfactants are activated by light whose intensity is 1-2 orders of magnitude lower than the laser intensities used in thermal-capillary actuation. We characterize the magnitude and rate of change in interfacial tensions of various fluid-fluid systems and demonstrate linear movement, merging, and complex 2D patterns of liquid droplets on various substrates. Furthermore, we present simple scaling arguments to explain the observed droplet-size-dependent velocity. The internal Marangoni flow is visualized via tracer particles and a simplified numerical model. The results demonstrated in this work open exciting doors for the use of photo-responsive surfactants for dynamic manipulation of multiphase fluid systems for energy, building, thermal management, and microfluidics applications.

\begin{figure*}[t!]
\centering
\includegraphics[width=\linewidth]{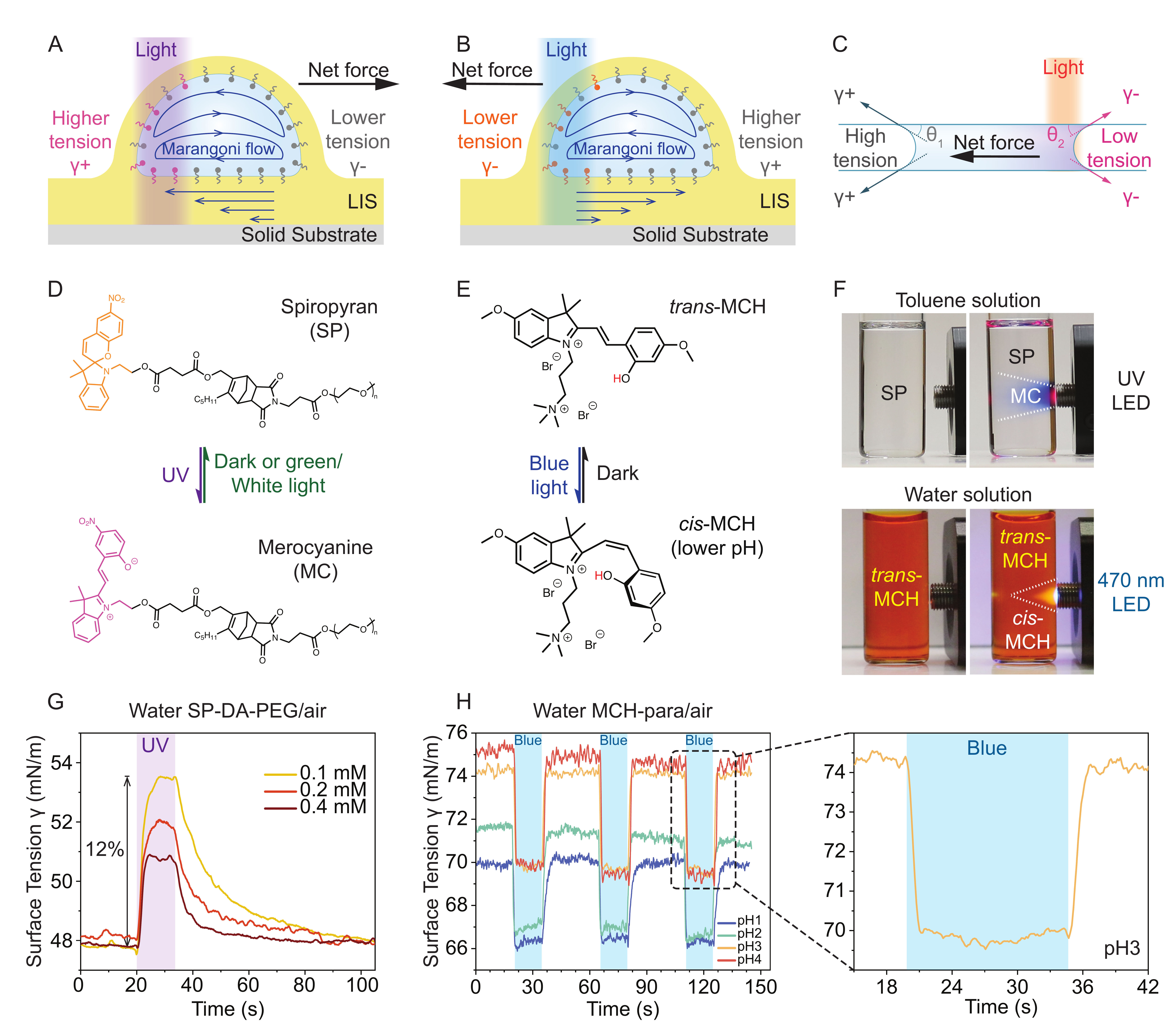}
\caption{\label{fig:1}Mechanisms of liquid movement driven by photo-responsive surfactants, molecular structures of the photo-responsive surfactants, and surface tension response. (A) Schematic of a lubricant-cloaked droplet on a liquid-infused surface (LIS). The interfacial tension increases under illumination, causing a Marangoni flow from the unilluminated region to the illuminated region, and a net shear force away from the light. (B) Schematic of a lubricant-cloaked droplet on LIS. The interfacial tension decreases under illumination, causing a Marangoni flow from the illuminated region to the unilluminated region, and a net shear force toward the light. (C)  Schematic of liquid in a microchannel or capillary tube. The surface tension changes under illumination, which results in an unbalanced total surface tension force on the liquid column and the subsequent liquid movement. Molecular structures and photoswitching of (D) SP-DA-PEG and (E) MCH-para. (F) The change of color of SP-DA-PEG in toluene under UV illumination and MCH-para in aqueous solution (phosphate buffer, pH 3) under 470 nm illumination. (G) Surface tension response of SP-DA-PEG in water under UV (365 nm) illumination with an optical intensity of 37.1 mW/cm\textsuperscript{2}. (H) Surface tension response of MCH-para in water at various pH levels under blue light (470 nm) illumination with an optical intensity of 31.8 mW/cm\textsuperscript{2}.}
\label{Fig. 1}
\end{figure*}

\section*{Results}

\subsection*{Working principle and characterization of surfactants} 
\medskip Photo-responsive surfactants or molecules are synthesized by attaching a photoswitchable molecule to a surfactant carrier. 
The photoswitchable moiety typically exists as a thermodynamically stable isomer in the dark.
Upon illumination with a specific wavelength, the isomer will undergo a conformational switch to a metastable isomer. 

When light is removed or a different wavelength of light is applied, the metastable isomer can switch back to the original thermodynamically stable isomer. 
This reversible isomer conversion can cause a change in the surface tension or interfacial tension of a solution containing the photo-responsive surfactant or molecule. We utilize this change in surface or interfacial tension to drive liquid motion on liquid-infused surfaces (LIS), inside microchannels with solid walls, and to form complex patterns on liquid surfaces. 

The working principle for driving fluid motion can be explained by two different mechanisms: interfacial shear caused by a photo-Marangoni (or chromo-capillary) flow and unbalanced surface tension forces. As illustrated in Fig. \ref{fig:1}A and \ref{fig:1}B, for a droplet on LIS consisting of a porous hydrophobic surface infused with lubricant oil, applying light on one side of the droplet will cause the photo-responsive surfactants or molecules under illumination to switch to the metastable isomer. This causes the local surface tension or interfacial tension $\gamma$ to increase (Fig. \ref{fig:1}A) or decrease (Fig. \ref{fig:1}B), which depends on the choice of surfactants and solvents. Meanwhile, the surfactants on the unilluminated side remain in the thermodynamically stable form. For droplets cloaked by the lubricant, an interfacial tension gradient will establish, generating a Marangoni flow from the low interfacial tension side to the high interfacial tension side. The interfacial flow generates a net shear force on the droplet, which causes it to move in the direction opposite to the interfacial flow direction. By adjusting the concentration of the surfactants, pH of the solution, and light intensity, the gradient in $\gamma$ can be quantitatively adjusted. On the other hand, for the liquid inside microfluidic channels with solid walls (Fig. \ref{fig:1}C), no Marangoni flow is present since there is no fluid-fluid interface. Rather, light causes the surface tension on the illuminated side of the liquid column to increase or decrease. The unbalanced surface tension force acting on the liquid along the channel direction becomes the main driving force.

To investigate photo-responsive surfactants with different characteristics, we selected two types of surfactants: SP-DA-PEG containing an electron-withdrawing nitro group (Fig. \ref{fig:1}D) which has been previously reported in our work \cite{zhao2022depinning}, and a newly synthesized merocyanine photoacid with two electron-donating methoxy groups incorporated at the \emph{para}-positions in the indolium and the phenolic moiety (MCH-para, Fig. \ref{fig:1}E). These molecules have different reaction timescales, are soluble in various types of solvents (aqueous and organic), and, as we show later, result in significant changes in the surface and interfacial tensions in the opposite directions due to differences in the photoswitching pathway. As shown in Fig. 1D, SP-DA-PEG can be tuned between neutral spiropyran (SP) and charged merocyanine (MC) forms upon 365 nm light irradiation; the reverse reaction occurs when the UV source is removed and can be accelerated by applying green light ($\sim$ 532 nm). In an aqueous solution, the photoswitch is accompanied by a change in color from transparent (SP form) to light pink (MC form) \cite{zhao2022depinning}. In toluene, the solution changes from transparent (SP form) to dark blue (MC form) (Fig. \ref{fig:1}F). 
The photoswitching mechanism of merocyanine photoacids is multi-step and depends on the equilibrium between the photoreactive states, solvent choice, pH of an aqueous solution, and substituent pattern \cite{johns2013physicochemical,satoh2011isomerization}. Recently, J. Beves et al. described a detailed switching mechanism of merocyanine photoacids \cite{wimberger2021large}. In the absence of light, merocyanine photoacids exist in three states: protonated merocyanine (MCH), deprotonated merocyanine (MC), and spiropyran (SP). At low pHs, MCH is the dominant form, while at basic conditions, the equilibrium shifts to MC which can undergo thermal ring-closure to form SP (SI Appendix, Fig. S7 A). For consistency, we investigated the aqueous solutions with a 1 to 4 pH range, where equilibrium completely shifted to the MCH form (SI Appendix, Fig. S7 B). At low pHs, the MCH-para switches instantaneously between \emph{trans}-MCH and the more acidic \emph{cis}-MCH form upon application and removal of blue light ($\sim$ 470~nm), respectively (Fig. \ref{fig:1}E and SI Appendix, Fig. S7 B). Such fast-switching kinetics is beneficial to result in a fast response of surface tension change, and maintain the surface tension gradient within a droplet. Fig. \ref{fig:1}F shows that 1 mM MCH-para aqueous solution bleaches from dark orange to yellow upon irradiation of 470~nm light, using a low-power fiber-coupled blue LED (5.6 mW/mm\textsuperscript{2}, Thorlabs M470F3). Moreover, we demonstrate that MCH-para has incredible hydrolytic stability at low pHs compared to previously reported merocyanine photoacids, which tend to hydrolyze due to the nucleophilic attack of water (SI Appendix, Fig. S6 A and B) \cite{hammarson2013characterization}. MCH-para was stable at pH 1-3 at room temperature for more than one month in aqueous media (SI Appendix, Fig. S7 C and D). This can be attributed to para- substitution with electron-donating methoxy groups both in indolium and phenolic moiety. Previously, Pezzato et al. reported that the para- substitution with electron-donating groups of either indolium or the chromene moiety increases hydrolytic stability \cite{berton2021light}. However, combinatory effects have not been previously investigated. 
We measured the surface tension responses of SP-DA-PEG and MCH-para aqueous solutions using a standard pendant drop method on a commercial tensiometer (Theta Flex, Biolin Scientific). Under light illumination, the surface tension of SP-DA-PEG increases (Fig. \ref{fig:1}G and SI Appendix, Fig. S11) while the surface tension of MCH-para solution decreases (Fig. \ref{fig:1}H). Fig. \ref{fig:1}G shows that the change in surface tension of the SP-DA-PEG aqueous solution can be tuned by varying the surfactant concentration. In this case, a 0.1~mM solution demonstrated a 5.6 mN/m change in surface tension within 8 s under 37.1 mW/cm\textsuperscript{2} illumination. Under the same illumination, increasing the concentrations to 0.2~mM and 0.4~mM reduced the response time to 7 s and 4~s, respectively, but also resulted in smaller surface tension changes (3.9 mN/m and 2.9 mN/m, respectively).  In addition, the surface tension response can also be tuned by the light intensity (SI Appendix, Fig. S11). The change in surface tension of a 0.2 mM SP-DA-PEG aqueous solution increases from 2.1 mN/m to 3.9~mN/m when the light intensity increases from 7.7 mW/cm\textsuperscript{2} to 37.1~mW/cm\textsuperscript{2}. These optical intensities are approximately 102 times lower compared to those used for thermocapillary actuation \cite{zhang2019light,gao2021optical}. The reverse switching was slower compared to forward switching. Compared to SP-DA-PEG, the surface tension response of MCH-para in water at various pH levels (phosphate buffer solution) is almost instantaneous (1.1 s for the forward reaction and 1.7~s for the reverse reaction), as illustrated in Fig. \ref{fig:1}H ($\sim$ 470~nm, 31.8 mW/cm\textsuperscript{2}).  We demonstrate multiple cycles of surface tension switching under repetitive pulsed blue light within a short period of time. The change in surface tension varies slightly from 3.6 mN/m at pH 1 to 5.7~mN/m at pH 4. 

\subsection*{Droplet movement on LIS} 
\begin{figure*}[h!]
\centering
\includegraphics[width=0.8\linewidth]{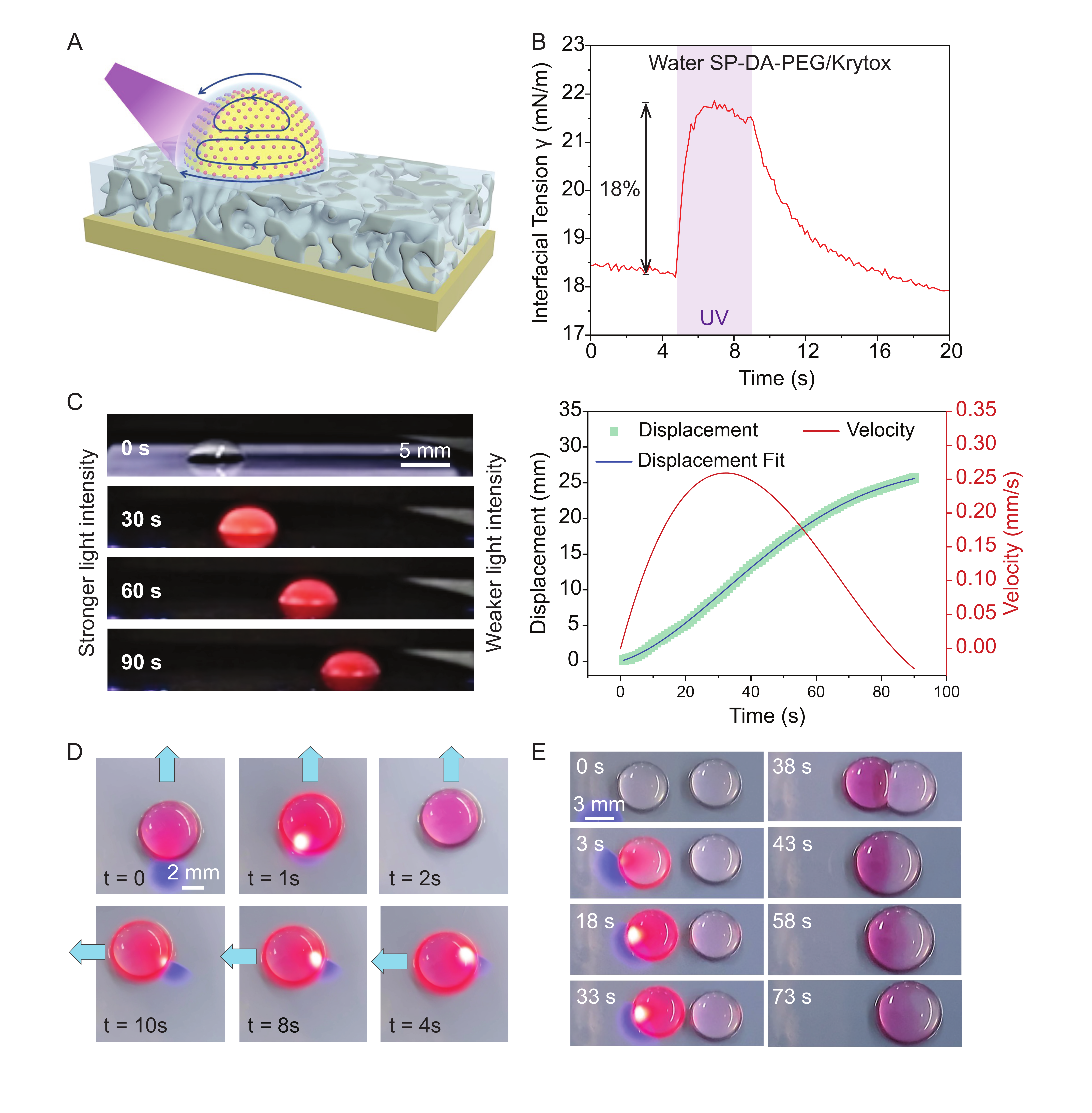}
\caption{\label{fig:2}Droplet movement (linear and 2D) and merging on liquid-infused surfaces (LIS) using SP-DA-PEG. (A) Schematic of a liquid droplet on LIS. (B) The interfacial tension response of water-Krytox interface with SP-DA-PEG when illuminated with 365 nm UV at 37.1 mW/cm\textsuperscript{2} optical intensity. (C) Time-lapse optical images (side view) of linear movement of a water droplet containing SP-DA-PEG on LIS directed by UV light with an intensity gradient. Light intensity is stronger on the left side and weaker on the right side. The line plot shows the displacement of the droplet and its velocity as functions of time. Time-lapse optical images (top-down view) of water droplets containing SP-DA-PEG on LIS (D) changing moving directions on demand, and (E) merging. }
\label{Fig. 2}
\end{figure*}

We first demonstrate and characterize droplet movement on a liquid-infused surface (LIS) using SP-DA-PEG and MCH-para (Fig. \ref{fig:2}A). LIS offers the advantages of high droplet mobility and low contact angle hysteresis, which make it particularly suitable as a microfluidic platform for fluid manipulation, as an anti-icing and anti-fouling surface, and as a condensation surface for power generation, thermal management, and desalination. The LIS used in this study consists of a porous PTFE membrane (SI Appendix, Fig. S12) infused with Krytox. As water droplets containing 0.2 mM SP-DA-PEG and 1~mM MCH-para (pH 3) respectively, were deposited on the LIS, we observed interference diffraction fringes on the surface of both droplets (SI Appendix, Fig. S13 A). This suggests that the droplets were cloaked by Krytox. In the absence of Krytox, the surface of the droplet shows no interference patterns (SI Appendix, Fig. S13 B).

Since both SP-DA-PEG and MCH-para are non-soluble in Krytox, they were anticipated to adsorb to the water-Krytox interface. For SP-DA-PEG, the water-Krytox interfacial tension was measured to increase from 18.2 mN/m to 21.8 mN/m ($\sim$20\% increase) after UV illumination (365 nm, 110~mW/cm\textsuperscript{2}), as shown in Fig. \ref{fig:2}B. The increase in interfacial tension is consistent with the increase of surface tension of water under UV (Fig. \ref{fig:1}G). Therefore, the application of UV light to one side of the droplet containing SP-DA-PEG will increase the interfacial tension locally. This generates a Marangoni flow from the unilluminated side to the illuminated side on both the top and bottom interfaces of the droplet, and two vortices form inside the droplet. A velocity gradient develops in the Krytox in LIS as a result of the interfacial flow, which exerts a net shear force on the droplet in the direction away from the light (Fig. \ref{fig:1}A).  

The mechanism described above for SP-DA-PEG is confirmed in Fig. \ref{fig:2}C. and Movie S1. When a water droplet containing SP-DA-PEG is placed in an environment where the light intensity is stronger on the left side and weaker on the right side realized using a continuously variable neutral density filter (Thorlabs NDL-10C-2) and a UV lamp (Analytik Jena, UVP BLAK-RAY B-100AP ALMP), the droplet moves toward the darker region (i.e., away from the light). Although the velocity is relatively slow ($\sim$250 $\mu$m/s, Fig. \ref{fig:2}C solid line plot), this is the first demonstration of droplet movement on LIS activated by photo-responsive surfactants. The output optical intensity after the UV light passed through the center of the filter was very low (6.1 mW/cm\textsuperscript{2}). In comparison, an equal-sized water droplet containing no surfactants remained stationary under the same illumination conditions (SI Appendix, Fig. S15), suggesting that droplet movement was not caused by the thermal-Marangoni effect from UV heating. Furthermore, the moving direction of the droplet can be controlled on the fly. As shown in Fig. \ref{fig:2}D and Movie S2, a water droplet was initially moving in the $+y$ direction driven by a UV laser (351.1~nm and 363.8 nm, beam diameter of 1.3 mm, optical power of 101 mW). Upon changing the laser position, the droplet immediately moved in the $-x$ direction. The photo-Marangoni effect can also achieve controlled merging of multiple droplets. As illustrated in Fig. \ref{fig:2}E and Movie S3, the same UV laser can drive a droplet to move toward and merge with another droplet. Since the left side of the as-formed droplet is rich in MC (pink), and the right side of the droplet contains only SP (transparent), a surface tension gradient remained in the merged droplet which caused it to move further even without UV light.

\begin{figure*}[h!]
\centering
\includegraphics[width=0.8\linewidth]{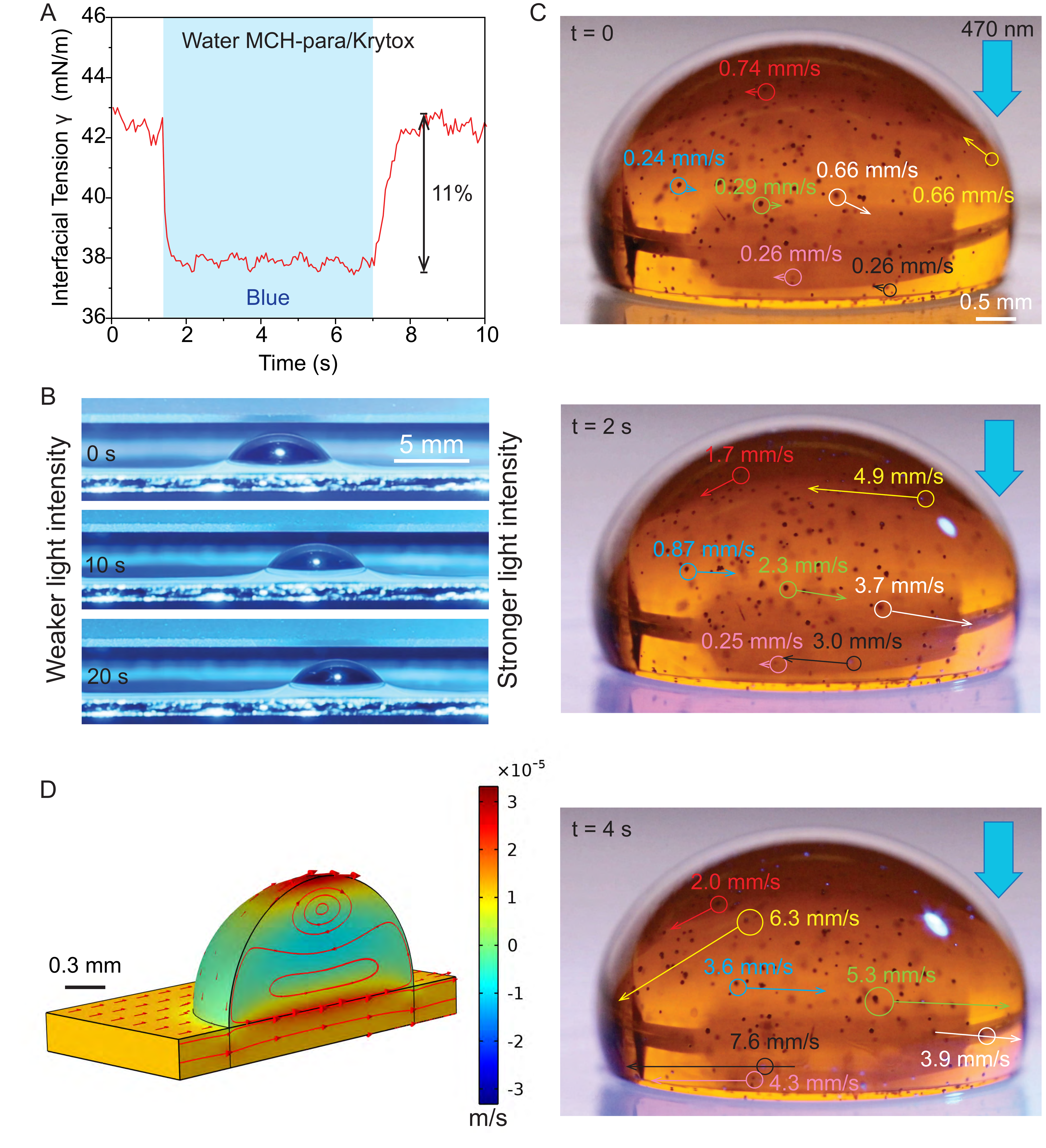}
\caption{Droplet movement on liquid-infused surfaces (LIS) using MCH-para. (A) The interfacial tension response of water-Krytox interface containing MCH-para when illuminated with 470 nm UV at 31.8 mW/cm\textsuperscript{2} optical intensity. (B) Time-lapse optical images (side view) of linear movement of a water droplet containing MCH-para on LIS directed by blue light with an intensity gradient. (C) Time-lapse optical images (side view) of the internal flow field indicated by tracer particles of a water droplet containing MCH-para. Blue light was illuminated on the right side of the droplet. The directions and lengths of arrows indicate the directions and magnitudes of the velocity of the tracer particles. (D) Computational results of light driven droplet motion on LIS by photo-Marangoni effect. }
\label{Fig. 3}
\end{figure*}

 We further investigated the movement and the internal flow field of water droplets (pH 3) containing 1 mM MCH-para on LIS. In contrast to SP-DA-PEG, the water-Krytox interfacial tension with MCH-para decreased from 42.9 mN/m to 37.9 mN/m (Fig. ~\ref{Fig. 3}A, 470 nm, 31.8 mW/cm\textsuperscript{2}), which is also consistent with surface tension measurement (Fig. \ref{fig:1}H). This reduction in the interfacial tension is also evident in the contact angle measurement (SI Appendix, Fig. S14), where the contact angle of an MCH-para water droplet on LIS reduced from 117$^{\circ}$ to 113$^{\circ}$ upon blue light (470~nm, 31.8~mW/cm\textsuperscript{2}) irradiation. Due to the opposite trend of interfacial tension change, we expect that the droplet containing MCH-para will move in the opposite direction, as shown in Fig. \ref{fig:1}B. As shown in Fig. ~\ref{Fig. 3}B, when the droplet is subject to a light intensity gradient (470 nm, stronger on the right side and weaker on the left side realized by the continuously variable neutral density filter, Thorlabs NDL-10C-2), the droplet moved toward the light in contrast to a droplet containing SP-DA-PEG moving away from the light. This indicates that we are able to precisely control the moving direction of the droplet by tuning the direction of surface tension response.

 We confirmed the internal flow field using tracer particles (Cospheric, sliver-coated hollow glass microsphere 25 - 65 $\mu$m, 0.9 g/cm\textsuperscript{3}). As shown in Fig. ~\ref{Fig. 3}C, when blue light (470 nm) illuminates the right side of the droplet, the local interfacial tension decreases. This causes a Marangoni flow as indicated by Fig. ~\ref{fig:1}B. The trajectory of tracer particles in Fig. ~\ref{Fig. 3}C and Movie S4 clearly showed that fluid indeed traveled from the low $\gamma$ side (illuminated region) to the high $\gamma$ side (un-illuminated region) along the water-Krytox interfaces. The directions and lengths of arrows on a few labeled tracer particles indicate the directions and magnitudes of the velocity of the tracer particles. We further developed a simplified numerical model to qualitatively illustrate the internal flow field using COMSOL Multiphysics. The model assumes that a shear stress is imposed on the top and bottom interfaces of a droplet on a thin lubricant layer. The flow is steady in a reference frame moving with the droplet. Fig.~\ref{Fig. 3}D shows two opposite-signed recirculation regions in the mid-plane of the droplet, which is consistent with the experimental observations in Fig.~\ref{Fig. 3}C. The detailed description of the numerical model is presented in the SI Appendix.

\subsection*{Droplet transport on liquids}
\begin{figure*}[h!]
\centering
\includegraphics[width=\linewidth]{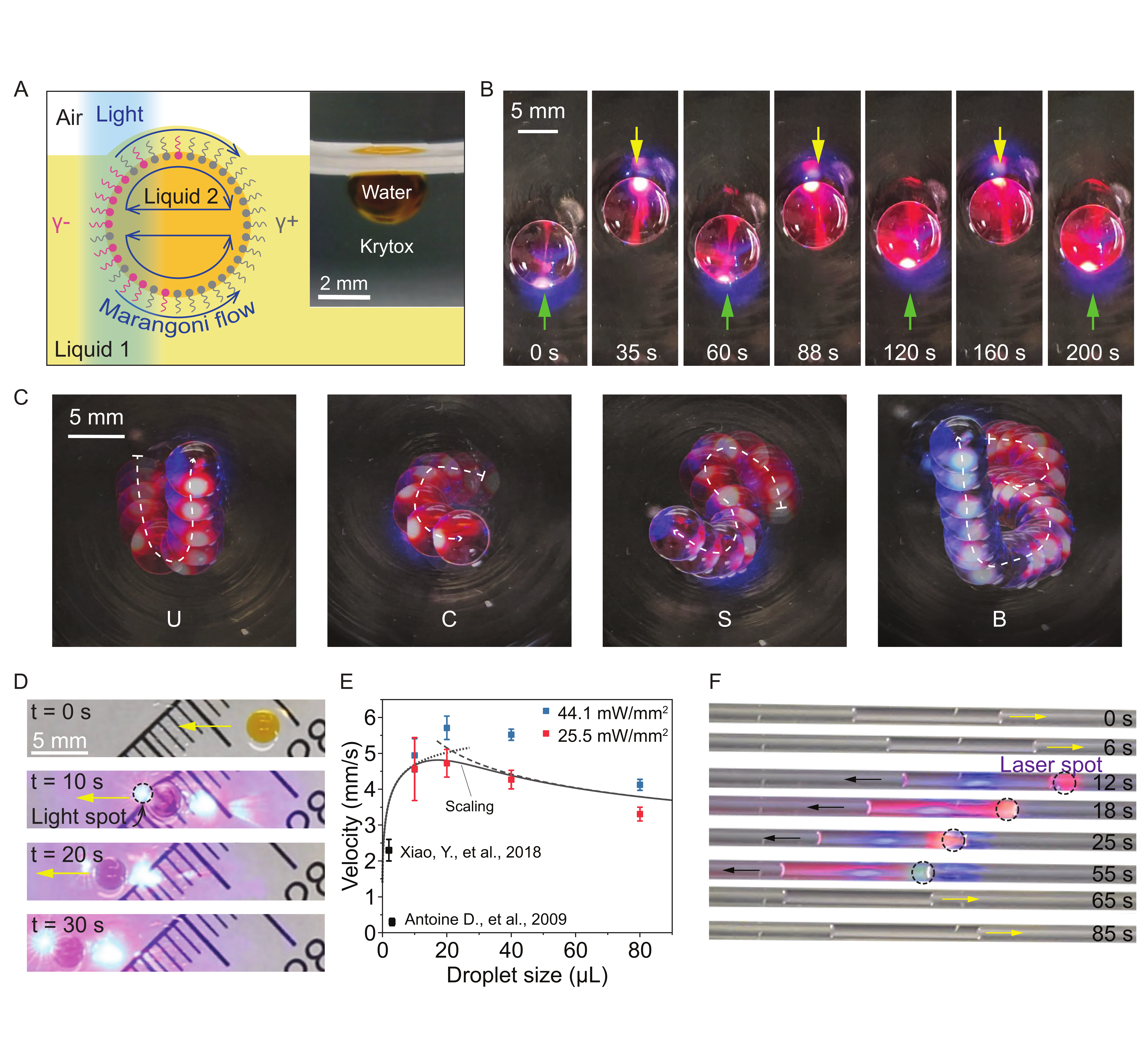}
\caption{Liquid movement (linear and 2D) on another immiscible liquid and inside solid-wall microchannels. (A) Schematic and optical image of a liquid droplet floating at the surface of an immiscible liquid. (B) Time-lapse optical images (top-down view) of linear movement and switching of the direction of a water droplet containing SP-DA-PEG on Krytox, directed by a UV laser. (C) Overlaid images (top-down view) of SP-DA-PEG droplet trajectories spelling the letters 'UCSB', driven by a UV laser. The time interval between images is 10 seconds. (D) Time-lapse optical images (top-down view) of linear movement of a water droplet containing MCH-para on Krytox, directed by a blue laser. (E) The average moving velocity of SP-DA-PEG droplets as a function of droplet volume and light intensity under UV illumination. Scalings for small and large drops are shown by dotted and dashed lines; a composite is shown by the continuous line. The results of \cite{xiao2018moving,diguet2009photomanipulation} are shown for reference. (F) Time-lapse optical images (side view) of a toluene liquid column containing SP-DA-PEG moving inside a glass capillary tube directed by a UV laser.}
\label{fig:dropOnLiquid}
\end{figure*}

We further investigate liquid motion directly on another immiscible liquid as a limiting case where there is no porous PTFE in the substrate (Fig.~\ref{fig:dropOnLiquid}A). Although linear droplet motions on a liquid substrate have been previously investigated by Xiao et al.~\cite{xiao2018moving} and Diguet et al.~\cite{diguet2009photomanipulation}, we demonstrate the ability to drive liquid in any arbitrary direction to form complex patterns and 2.5 times higher maximum droplet velocity. Water droplets containing 0.2 mM of SP-DA-PEG were placed on a thick layer ($>$ 1 cm) of higher-density Krytox lubricant. As shown in Fig.~\ref{fig:dropOnLiquid}B and Movie~S5, the application of a UV laser (351.1~nm and 363.8 nm, beam diameter of 1.3 mm, optical power of 101 mW) can actuate droplet motion and switch its moving direction repeatedly. Moreover, we show in Fig.~\ref{fig:dropOnLiquid}C and Movie~S6 that the same laser can precisely drive droplets to describe a complex pattern, shown here by spelling the letters `UCSB'. Fig.~\ref{fig:dropOnLiquid}C comprises overlaid time-lapse images to illustrate the trajectory of the droplet. Consistently with our previous experiments, droplets containing SP-DA-PEG move away from the light, as shown in Figs.~\ref{fig:dropOnLiquid}B-C, whereas droplets containing MCH-para move towards the light (450~nm, 354\,mW/cm\textsuperscript{2}), as indicated in Fig.~\ref{fig:dropOnLiquid}D. 

Furthermore, we studied how velocity depends on size and light intensity. Fig.~\ref{fig:dropOnLiquid}E shows the average moving velocity of SP-DA-PEG droplets as a function of droplet volume and light intensity. The maximum velocity was observed when the droplet volume was approximately 20\,$\mu$L, which is 2.5 times higher compared to previous studies~\cite{xiao2018moving,diguet2009photomanipulation}. Velocity decreases gradually as drop volume is increased past 20\,$\mu$L.

We suggest simple scaling arguments to explain the observed maximum in drop velocity. 
                     Within the drop, interfacial tension increases rapidly for fluid within the laser spot; as the fluid moves away from the light, its interfacial tension gradually returns towards its `dark' value (as shown earlier in Fig.~\ref{Fig. 2}B). For smaller volumes, fluid takes less time to travel the drop length, leading to an incomplete reverse photo-reaction and therefore a smaller decrease in interfacial tension. As a result, the interfacial tension difference across the drop becomes weaker. Balancing this Marangoni force and the fluid resistance (see SI Scaling theory) we obtain the dotted curve for the velocity $V$ in Fig.~\ref{fig:dropOnLiquid}E. For small drops 
\begin{equation}
V \sim \sqrt{\frac{\Delta \gamma}{\mu} \frac{2R-d}{t_\mathrm{rev}}} \quad \quad \text{if}  \quad \quad 2R-d \ll V t_\mathrm{rev} 
\end{equation}
where $\mu$ is the viscosity of Krytox, $R$ the drop radius, $d$ the beam diameter, $t_\mathrm{rev}$ the reverse reaction timescale, and $\Delta \gamma$ the maximum surface tension change achievable under the given illumination (see Supplementary Material). For larger drops, there are several contributing factors that could explain the decrease in velocity with increasing volume. We note for example that the largest drops have Reynolds numbers $Re = \frac{\rho V 2R}{\mu} \approx 4$, such that some of the Marangoni force is expended to overcome the inertia of the surrounding fluid, thereby decreasing propulsive efficiency. A scaling based on this hypothesis is shown by the dashed line in Fig.~\ref{fig:dropOnLiquid}E, which approximately follows $V\sim R^{-1/2}$ (see SI Scaling theory); the continuous line is a composite of the results for small and large drops.


Previous studies reported that the photo-Marangoni effect occurs in isothermal conditions (i.e., without localized heating). However, we note that heating from light absorption is possible \cite{Seshadri2020-be}. 
Here, we show that under the experimental conditions investigated in our study, the photo-Marangoni effect is the dominant factor despite the co-existence of a thermal-Marangoni effect. For SP-DA-PEG, illumination causes the surface tension and interfacial tension to increase, while increased temperature causes a decrease. The direction of fluid motion observed in our experiments confirms that the contribution of temperature change must be small compared to the photo-Marangoni stress. On the other hand, for MCH-para solutions, since both heating and photo-switching decrease the surface and interfacial tension, we measured the change of surface tension caused purely by an increase in temperature from 20$^{\circ}$C to 60$^{\circ}$C using a custom syringe heating system (SI Appendix, Fig.~S8 A). Surface tension decreases with temperature following the equation
\begin{equation}
\centering
 \gamma \left[\mathrm{\frac{mN}{m}}\right] = -0.26\; T [^{\circ}C] + 76.66
\label{eqn:temp}
\end{equation}
To estimate the separate contributions from heating and from photoswitching on the change of surface tension shown in Fig.~\ref{Fig. 1}H, we measured the temperature rise of a water droplet (pH 3) containing MCH-para under the same illumination condition (470 nm 31.8 mW/cm\textsuperscript{2}) using an infrared camera  (Telops, M3K, SI Appendix, Fig.~S9\,A). We used the emissivity of pure water ($\epsilon$ = 1) to approximate the emissivity of the MCH-para aqueous solution (see SI Appendix, Fig.~S8\,C for details). As shown in SI Appendix, Fig.~S9\,C and D, the highest temperature within the MCH-para droplet increased from 19.6$^{\circ}$C to 20.6$^{\circ}$C after 150\,s of blue light illumination (31.8 mW/cm\textsuperscript{2}). According to equation~\ref{eqn:temp}, this temperature change only generates a 0.26 mN/m decrease in surface tension, which is only 5.9\% of the surface tension change in Fig.~\ref{Fig. 1}H. Furthermore, the temperature increase is negligible within the timescale of the photoswitch ($< 2$s). Therefore, under these optical intensities, even though both heating and photo-switching contribute to a decrease in the surface tension, the photoisomerization-induced Marangoni effect is the primary factor in driving droplet motion. When applying a higher optical intensity, the surface tension changes caused by heating and photoswitching both increase in their magnitude; their relative contributions can be estimated by the method presented here.

\subsection*{Liquid transport in microchannels}
In addition to modulating droplet movement on LIS and liquid surfaces, here we also demonstrate the direct movement of liquid on solid walls of microchannels or capillary tubes using light-responsive surfactants. Fig. ~\ref{fig:dropOnLiquid}F shows time-lapse side  view images of a glass capillary tube (0.5 mm diameter) containing toluene with 0.1 mM SP-DA-PEG surfactants (Movie S7).  The glass tube was intentionally tilted so that under the influence of gravity, the liquid naturally slide toward the right (from 0 s to 6 s). Application of light (351.1 nm and 363.8 nm, beam diameter of 1.3 mm, optical power of 101~mW) on the right side of the liquid column caused the liquid to move toward the left (12 s to 55~s). When the light was removed, the liquid resumed moving towards the right due to gravity (65 s to 85 s). In our previous work, the surface tension of toluene containing SP-DA-PEG was measured to decrease with UV illumination \cite{zhao2022depinning}. Since the only fluid-fluid interface is the toluene-air interface at the two ends of the liquid column, the mechanism driving fluid motion is not the Marangoni effect, but rather, the unbalanced surface tension forces on the liquid. Specifically, the surface tension on the left unilluminated side remains the same whereas the surface tension on the right illuminated side reduces, contributing to a net force driving the liquid to move away towards the higher surface tension side. In addition, a change in contact angle can also contribute to the horizontal component of the net surface tension force (Fig. \ref{fig:1}C). However, we did not observe a clear change in the contact angle with and without illumination. 

\section*{Discussion}
We demonstrated a novel approach for dynamic manipulation of droplet motions on lubricant-infused surfaces (LIS), solid-wall microchannels, and on liquid substrates using photo-responsive surfactants designed and synthesized by us. The surfactants SP-DA-PEG and MCH-para selected in this study have different reaction timescales (1 to 10 seconds) and cause surface tension and interfacial tension to change in the opposite direction. MCH-para is particularly stable in low-pH aqueous solutions. In addition, the photo-switch is activated with UV or blue light with optical intensity 1-2 orders of magnitude lower than what is typically used by thermal-capillary actuation. We show for the first time that droplets are able to move and merge in linear motions on LIS, and their moving directions can be changed dynamically by light. The Marangoni-induced internal fluid flow was also visualized by analyzing the trajectory and velocity of tracer particles, which confirmed our proposed mechanism. In particular, the moving direction depends on if surfactants increase or decrease the surface or interfacial tension. Furthermore, we show fast motion (up to 5.8 mm) of droplets floating on another liquid substrate, and that droplets are able to move in arbitrary directions on a 2D surface to form complex patterns. We developed scaling arguments to explain the observed droplet-size dependent velocity and a simpliﬁed numerical model to qualitatively illustrate the internal ﬂow.

In addition to the photo-Marangoni effect, we also demonstrate the movement of liquid inside solid-wall microchannels which is caused by unbalanced surface tension along the moving direction. This can potentially be used as a non-contact pumping mechanism for microfluidic applications. 

The droplet manipulation platform presented in this work opens doors for applications requiring non-contact, non-invasive, dynamically tunable and programmable fluid motions with minimal external energy input. Compare with digital microfluidics realized using electrowetting and surfactant-mediated electro-dewetting, our approach does not require any fabrication of microelectrodes and electrical connections. Light can also be focused down to the diffraction-limit length, potentially enabling microfluidic manipulation down to the 1-micron spatial resolution. This however requires further development of photo-responsive surfactants with faster reaction kinetics. In addition, photo-surfactants that can be soluble in the lubricant of LIS can help with droplet removal in condensation processes, which can potentially enhance condensation heat transfer significantly for power generation and desalination applications. The approach presented here also serves as a powerful method for multi-phase processes in microgravity, where droplet motions cannot be achieved with gravity.

\section*{Materials and Methods}
\subsection*{Sample preparations}
The photoactive surfactant SP-DA-PEG used in this study is soluble in toluene and water. The solutions with the desired concentration were prepared by sonicating appropriate amount of SP-DA-PEG in solvents (35). The photoactive surfactant MCH-para used in this study has a high solubility in aqueous solutions. A concentrated stock solution was used to create a 1 mM MCH-para solution. The stock solution of MCH-para (C = 2 mM) was prepared by adding 15 mg (Molecular Weight = 584.38 g/mol, 0.026 mmol) in 12.83 mL of deionized water in a 20 ml vial, wrapped in aluminum foil and stored at 4 $^{\circ}$C. Final solutions were prepared in a 4 mL cuvette by adding 0.6 mL of DI water followed by 0.4~mL of corresponding buffer solution (pH 1-4) and equilibrated for 10 minutes. After which, 1 mL of MCH-para stock solution was added and equilibrated for 15 minutes in the dark.

\subsection*{Materials}
Surfactants, SP-DA-PEG and MCH-para, were synthesized in the laboratory (SI Appendix, Note S1). Buffer solutions were prepared in the laboratory by adding NaOH or HCl to Na\textsubscript{2}HPO\textsubscript{4} and NaH\textsubscript{2}PO\textsubscript{4} solutions (reagents were obtained from Sigma Aldrich, Oakwood Chemical, or Fisher Scientific). Solvents, deionized water and toluene, were purchased from Sigma-Aldrich. Lubricant oil, Krytox GPL 100, was supplied by Miller-Stephenson. PTFE thin films were obtained from Sterlitech (PTFE unlaminated membrane filters, 0.2 micron, 90 mm; the typical porosity of the PTFE is 90\%). SiO\textsubscript{2} microsepheric tracer particles were obtained from Cospheric (sliver-coated hollow glass microsphere 25 - 65 $\mu$m, 0.9 g/cm\textsuperscript{3}). All reagents were of analytical grade and used as received. 

\subsection*{Light sources}
A custom multiline argon ion UV laser (351.1 \& 363.8 nm) with a beam diameter of 1.3 mm and maximum light intensity of 44 mW/mm\textsuperscript{2} was used for SP-DA-PEG droplet movement experiments. The fiber-coupled 365 nm UV LED (M365FP1, 9.8 mW Min Fiber-Coupled LED), 470~nm blue LEDs (M470F3, 17.2 mW Min Fiber-Coupled LED and M470L3-C1, 350 mW Collimated LED), 530 nm green LED (M530F2, 6.8 mW Min Fiber-Coupled LED) and white LED (MCWHF2, 21.5 mW Min Fiber-Coupled LED) were directly bought from Thorlabs, Inc. The blue laser was built with complete laser diode (LD) operation starter sets from Thorlabs, Inc, including a laser diode controller (LDC 200C), a thermoelectric temperature controller (TED 200C), and a 450 nm blue laser diode (L450P1600MM 1600 mW, 5.6~mm, G Pin Code, MM). A high-intensity UV lamp (216 mm x 140 mm, UVP High Intensity Lamp, Analytik Jena) coupled with an ND filter (Thorlabs NDL-10C-2) was used to conduct experiments on driving SP-DA-PEG droplet motion. 

\subsection*{Instrument and Characterization}
Surface tension, interfacial tension and contact angle were characterized using a standard pendant drop method on commercial tensiometer, Biolin Scientific, Theta Flex. Optical power was measured by using optical power meter (Newport, 843-R) connected with a thermopile Sensor (Newport 919P-003-10, 3 W, 10 mm, 0.19-11~$\mu$m). Optical images and videos were taken by a Canon EOS 80D camera. IR images were recorded by a high-performance Telops M3K infrared camera. UV–Vis absorption spectra were recorded on an Agilent 8453 UV–Vis spectrometer from 200 to 1200 nm wavelengths. \textsuperscript{1}H and \textsuperscript{13}C NMR spectra were recorded on a Bruker 500 MHz NMR spectrometer. The pH of the solutions was measured by using a Thermo Scientific Orion Star™ A111 Benchtop pH Meter. The scanning electron microscope (SEM) image was obtained from a FEI Sirion SEM. 

\subsection*{Fabrication of the lubricant-infused surface (LIS)}
A porous thin film PTFE was obtained from Sterlitech. PTFE film is cut into pieces and adhered to a flat surface, such as a silicon wafer or a microscopy slide. Extra ethanol was applied to the PTFE and allowed to completely infiltrate before evaporating at room temperature in the fume hood. To create the lubricant-infused surface (LIS), inject Krytox GPL 100 lubricant oil into the PTFE that was adhered to the solid, flat surface. 

\subsection*{Data Availability}
The data described in this article are available in figshare at \\10.6084/m9.figshare.22732736 
and 10.6084/m9.figshare.21866181

\section*{Acknowledgements}Research reported in this publication was supported by the Center for the Advancement of Science in Space, the National Science Foundation, and sponsored by the International Space Station National Laboratory under grant number 2025655. The NMR results reported here benefited from the shared analytical instrumentation of the Spectroscopy Facility at the UCSB Chemistry and Biochemistry Department. We thank Vijay Kumar from UCSB for assistance in the construction of the custom syringe heating system and IR temperature measurements.

\bibliographystyle{unsrt}
\bibliography{sample}

\newpage
\section*{Supporting Information}
\subsection*{Synthesis and Characterization of SP-DA-PEG and MCH-para}
\subsubsection*{Materials and Instrumentation}\mbox{} \\
All reagents were obtained from Sigma Aldrich, Oakwood Chemical, or Fisher Scientific and were used as received without further purification unless specified. Acetonitrile and dichloromethane were dispensed from a solvent purification system immediately before use or stored over 3Å molecular sieves. Ethanol and chloroform were stored over 3Å molecular sieves before use. Analytical thin-layer chromatography (TLC) was performed with Merck silica gel 60 F254 pre-coated glass plates and visualized by UV light or stained with p-anisaldehyde. \textsuperscript{1}H and \textsuperscript{13}C NMR spectra were recorded on a Bruker 500 MHz NMR spectrometer. Chemical shifts are reported relative to residual solvent peaks ($\delta$ 7.26 ppm for CDCl3, 2.50 ppm for DMSO-\emph{d\textsubscript{6}}:) in \textsuperscript{1}H NMR, and $\delta$ 77.20 ppm for CDCl\textsubscript{3}, 39.52 ppm for DMSO-\emph{d\textsubscript{6}} in \textsuperscript{13}C NMR). 

\subsubsection*{UV–Vis spectroscopy and UV–Vis kinetic measurements}\mbox{} \\
UV–Vis absorption spectra were recorded on an Agilent 8453 UV–Vis spectrometer from 200 to 1200~nm wavelengths.
The photoinduced optical absorption kinetics of SP-DA-PEG and MCH-para were measured on a custom pump-probe setup. The pump beam was generated by a high-power LED (light-emitting diode) source (Thorlabs) coupled into a multimode optical fiber terminated with an output collimator. The UV pump source was generated by the fiber-coupled UV LED (Thorlabs M365FP1, 365 nm, 9.8 mW Min Fiber-Coupled LED); the green pump source was produced by green LED (Thorlabs M530F2, 530 nm 6.8~mW Min Fiber-Coupled LED) and the blue pump source by blue LED (M470F3, 17.2 mW Min Fiber-Coupled LED), which was positioned to illuminate the sample directly. The probe beam was generated by a High-Power MINI Deuterium Tungsten Halogen Source (Ocean Optics DH-MINI) w/shutter 200-2000 nm coupled into a multimode optical fiber terminated with an output collimator. The probe light was altered by a shutter (Uniblitz CS25) and controlled manually or through a digital output port (National Instruments USB-6009) using the LabVIEW program. A 10x10 mm\textsuperscript{2} rectangular spectrophotometer cell was placed in a sample holder. The solution was continuously stirred by a miniature stirring plate inserted into the sample holder (Starna Cells SCS 1.11). A system of lenses directed the probe beam into the detector (Ocean Optics Flame-S1-XR spectrometer), which acquired the spectra of the probe light. The detector was connected to a PC via a USB port. The experiment was controlled by a National Instrument LabVIEW program which collected the probe light spectra, determined sample optical absorption spectra, and controlled pump and probe light sources. Kinetics experiments with SP-DA-PEG were performed in 0.05 mM solutions (SI Appendix, Fig. S1 A and B). 

\begin{figure*}[h]
\centering
\includegraphics[width=\linewidth]{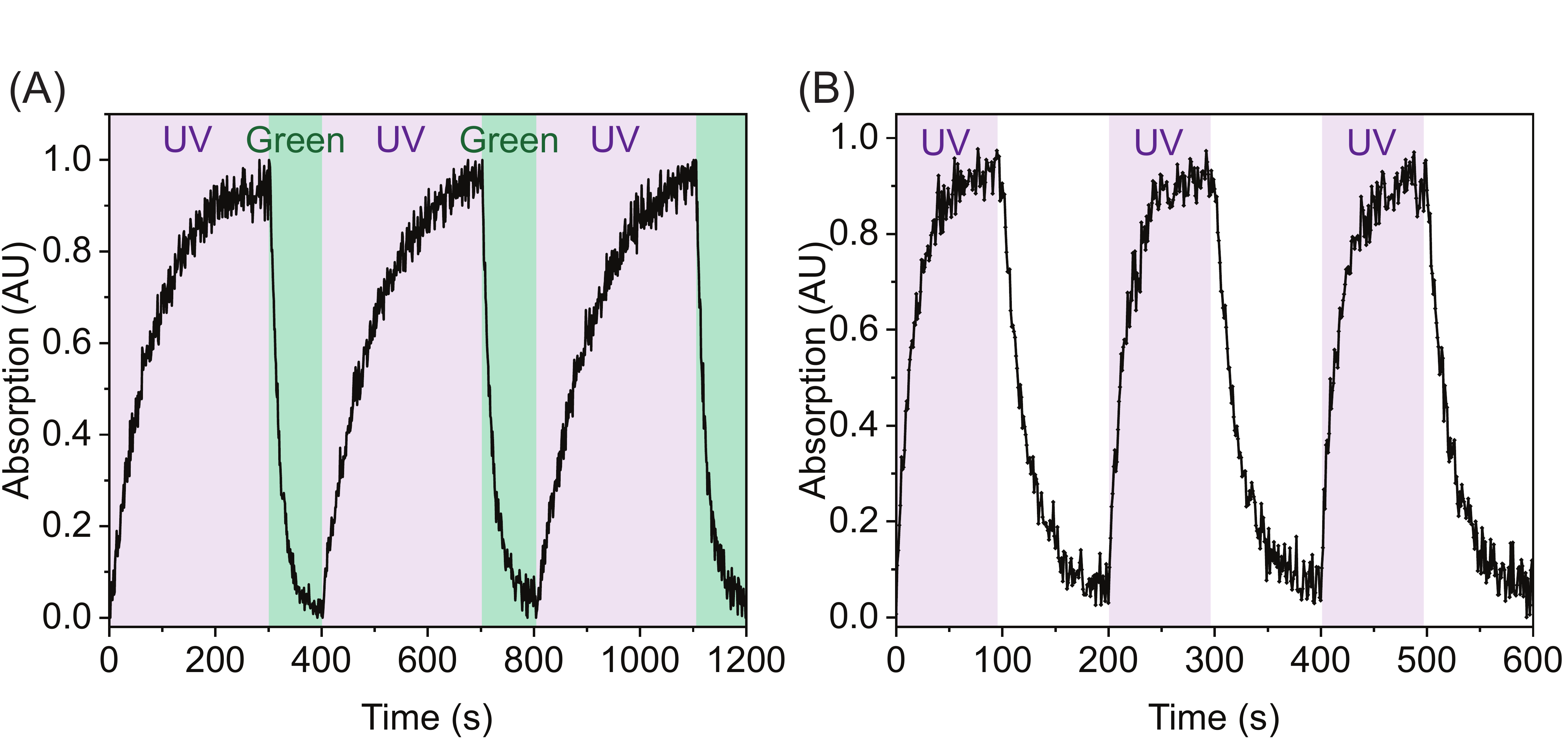}
{Figure S1: Pump probe kinetics measurements of SP-DA-PEG in (A) water with alternative 365 nm and 530 nm lights and (B) toluene irradiated with 365 nm light.}
\label{Fig. S1}
\end{figure*}

\subsubsection*{Solution preparation for the experiments and UV-Vis studies}\mbox{} \\
Solution preparation procedures were modified from Wimberger et al \cite{wimberger2021large}.
\paragraph{1. Monobasic buffer preparation:}
Buffer solutions ranging from pH 1-10 were prepared by adding NaOH (1 M) or HCl (1 M) to Na\textsubscript{2}HPO\textsubscript{4} and NaH\textsubscript{2}PO\textsubscript{4} solutions (100 mM). DI water was used for all solutions and buffers. The pH of the final solutions was measured by Thermo Scientific Orion Star™ A111 Benchtop pH Meter. Prepared buffers were used for both UV-Vis studies and experiments.
\paragraph{2. The following procedure was used for experiments:}
The stock solution of MCH-para (C = 2 mM) was prepared by adding 15 mg (Molecular Weight = 584.38 g/mol, 0.026 mmol) in 12.83 mL of DI water in 20 mL vial, wrapped in aluminum foil and stored at 4 $^{\circ}$C. Final solutions were prepared in a 2 mL cuvette by adding 0.6 mL of DI water followed by 0.4 mL of corresponding buffer solution (pH 1-4) and equilibrated for 10 minutes. After which, 1 mL of MCH-para stock solution was added and equilibrated for 15 minutes in the dark.
\paragraph{3. UV-Vis samples were prepared as follows:}
The stock solution of the MCH-para (C = 0.25 mM) was prepared by adding 14.6~mg (Molecular Weight = 584.38 g/mol, 0.025 mmol) in 100~mL of DI water into a volumetric flask wrapped in aluminum foil and stored at 4 $^{\circ}$C. Solutions for UV-Vis analysis were prepared in a 1 cm cuvette by adding 0.57 mL of the corresponding phosphate buffer (pH 1-10) followed by 2.052 mL of DI water, placed in the spectrophotometer, and allowed to equilibrate thermally for 10 minutes. Then 0.228 mL of MCH-para stock solution was added to the final concentration of C = 0.020 mM, with continuous stirring of the solution. Scans were collected after 15 minutes of equilibration in the dark.
\subsubsection*{Synthesis}
\paragraph{1. Synthesis of SP-DA-PEG:}\mbox{}\\

\begin{figure*}[h]
\centering
\includegraphics[width=\linewidth]{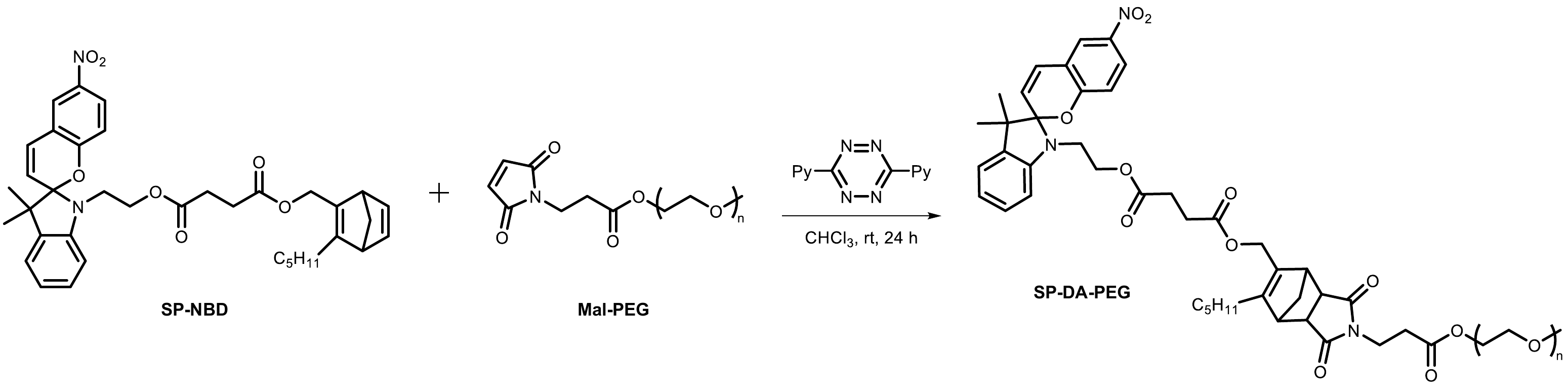}
\end{figure*}

\textbf{SP-DA-PEG} and its precursors \textbf{SP-NBD} and \textbf{Mal-PEG} were synthesized as previously reported by Seshadri et al. \cite{seshadri2021influence}. and kept under a nitrogen atmosphere at 0 °C. Spectral data matches those reported in the literature \cite{seshadri2021influence,zhao2022depinning}.\\
\textbf{\textsuperscript{1}H NMR} (500 MHz, CDCl\textsubscript{3}) $\delta$ 8.02–8.00 (m, 2H), 7.20 (t, 1H), 7.08 (d, 1H), 6.93 (dd, 1H), 6.89 (t, 1H), 6.75 (d, 1H), 6.67(d, 1H), 5.89 (d, 1H), 4.68 (dd, 1H), 4.30 (dd, 1H) 4.27–4.19 (m, 4H), 3.64 (m, 47H), 3.55–3.53 (m, 2H), 3.38 (s, 3H), 2.63–2.54 (m, 4H), 2.48–2.45 (m, 2H), 2.19–2.14 (m, 1H), 1.82–1.77 (m, 1H), 1.71 (d, 1H), 1.50–1.47 (m, 2H) 1.37–1.18 (m, 9H), 1.17 (s, 3H), 0.87 (t, 3H) ppm. M\textsubscript{n}(NMR) = 1390 g/mol.

\paragraph{2. Synthesis of merocyanine MCH-para:} \mbox{}\\

\begin{figure*}[h]
\centering
\includegraphics[width=\linewidth]{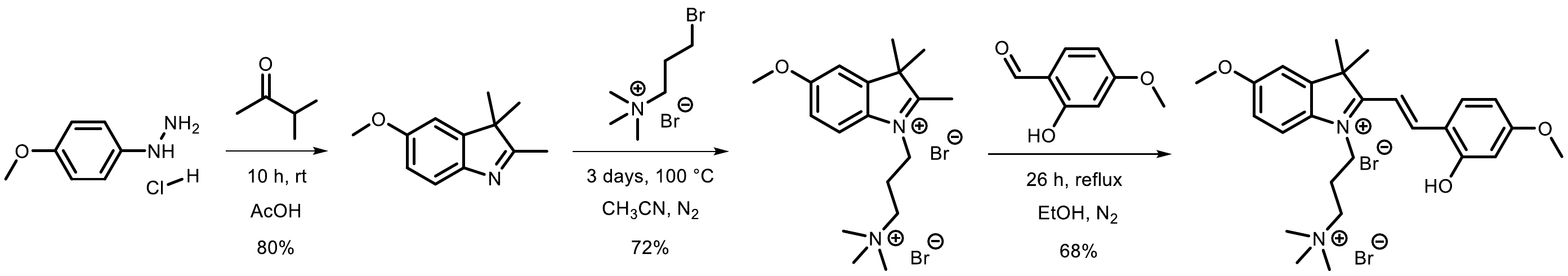}
\end{figure*}

\subparagraph{5-Methoxy-2,3,3-trimethyl-3H-indole}\mbox{}\\

\begin{figure*}[h]
\centering
\includegraphics[width=6.8cm]{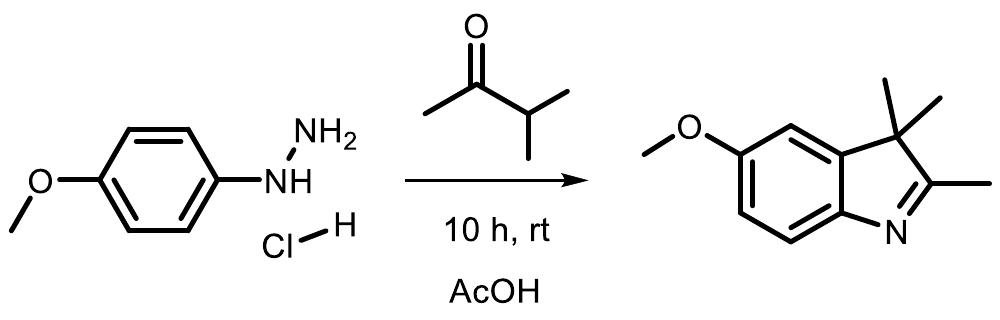}
\end{figure*}

Synthesis of 5-Methoxy-2,3,3-trimethyl-3H-indole was adapted from Wimberger et al \cite{wimberger2021large}.
To a solution of 3-methyl-2-butanone (2.48 mL, 2.0 g, 23.2 mmol, 2.00 eq.) in 50 mL of glacial acetic acid was added 4-methoxyphenylhydrazine hydrochloride (2.0 g, 11.6 mmol, 1.00 eq.). The mixture was heated to 140 $^{\circ}$C for 20 min to dissolve 4-methoxyphenylhydrazine hydrochloride completely. The purple solution was stirred at room temperature overnight, after which potassium hydroxide pellets were added slowly to neutralize the mixture. The solution was extracted with diethyl ether (3 × 50~mL), and the combined organic phases were washed with brine (35 mL), dried over magnesium sulfate, and concentrated under reduced pressure. The final product can be purified by column chromatography (hex/EtOAc = 7/3 → 1/1) if necessary. 1.76 g (80\% yield). Due to possible decomposition at room temperature, the product was kept in the fridge.\\
\textbf{\textsuperscript{1}H NMR} (500 MHz, CDCl\textsubscript{3}) $\delta$ 7.43 (dd, \emph{J} = 8.2, 0.7 Hz, 1H), 6.87 – 6.78 (m, 2H), 3.83 (s, 3H), 2.24 (s, 3H), 1.28 (s, 6H) ppm.\\
Spectral data matches the reported literature \cite{wimberger2021large}.
\begin{figure*}[h]
\centering
\includegraphics[width=15cm]{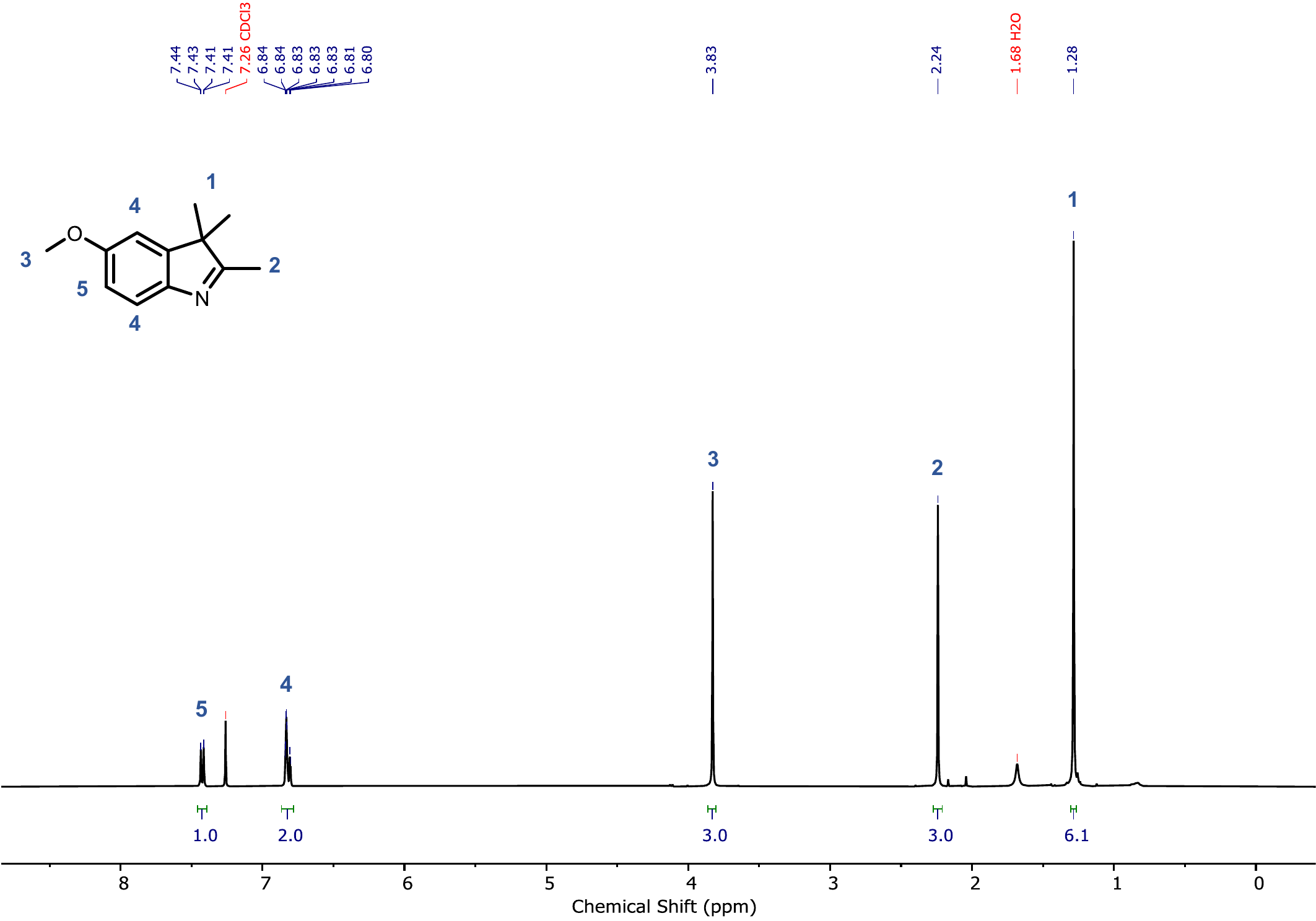}
Figure S2: {\textsuperscript{1}H NMR spectrum of 5-Methoxy-2,3,3-trimethyl-3H-indole.}
\label{Fig. S2}
\end{figure*}

\subparagraph{5-Methoxy-2,3,3-trimethyl-1-(3-(trimethylammonio)propyl)-3H-indol-1-ium dibromide}\mbox{}\\
\begin{figure*}[h]
\centering
\includegraphics[width=7.7cm]{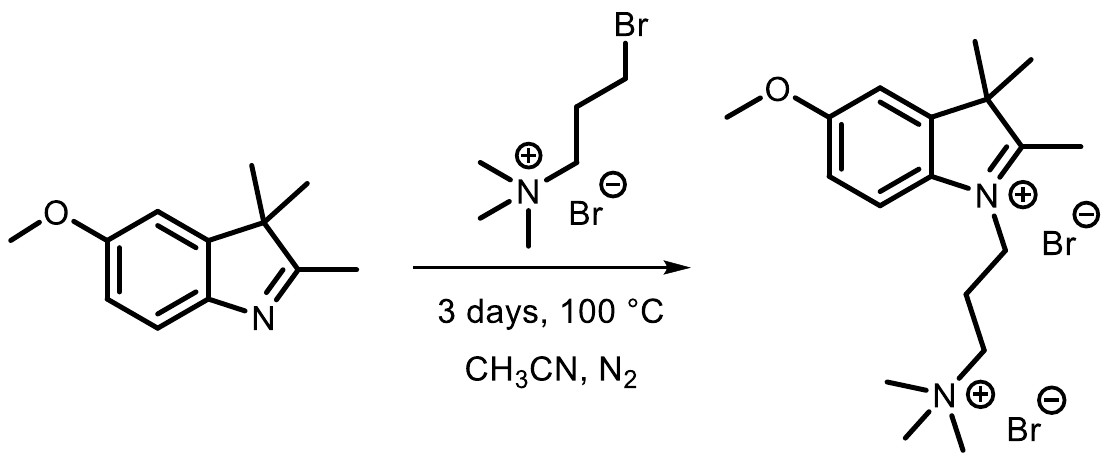}
\end{figure*}

Synthesis of 5-Methoxy-2,3,3-trimethyl-1-(3-(trimethylammonio)propyl)-3H-indol-1-ium dibromide was adapted from Wimberger et al \cite{wimberger2021large}.
To an oven-dry round bottom flask equipped with a magnetic stir bar, a condenser was added (3-bromopropyl)trimethyl-ammonium dibromide (1.5 g, 5.75 mmol, 1.10 eq) followed by a 5-methoxy-2,3,3-trimethyl-3H-indole (0.99 g, 5.22 mmol, 1.00 eq.) in deoxygenated absolute acetonitrile (25 mL). The reaction mixture was heated to 100 $^{\circ}$C and refluxed under a nitrogen atmosphere for three days. After cooling down to room temperature, acetonitrile was removed under reduced pressure. The resulting purple solid was dissolved in 70 ml of water and washed with diethyl ether (3 × 40 mL) to remove impurities. The aqueous phase was evaporated in a vacuum, and the product was dissolved in a minimum amount of methanol and precipitated in cold diethyl ether, filtered, and dried in a vacuum oven overnight.
The final product contains a small amount of (3-bromopropyl)trimethyl-ammonium dibromide, which due to similar solubility, was impossible to remove. Therefore, it was used without further purification. 1.69 g (72\% yield).\\
\textbf{\textsuperscript{1}H NMR} (500 MHz, DMSO-\emph{d\textsubscript{6}}:) $\delta$ 7.99 (d, \emph{J} = 8.9 Hz, 1H), 7.51 (d, \emph{J} = 2.5 Hz, 1H), 7.17 (dd, \emph{J} = 8.9, 2.5 Hz, 1H), 4.46 (t, \emph{J} = 8.0 Hz, 2H), 3.87 (s, 3H), 3.60 – 3.57 (m, 2H), 3.11 (s, 9H), 2.85 (s, 3H), 2.32 (qt, \emph{J} = 7.9, 4.9 Hz, 2H), 1.54 (s, 6H) ppm.
\begin{figure*}[h]
\centering
\includegraphics[width=15cm]{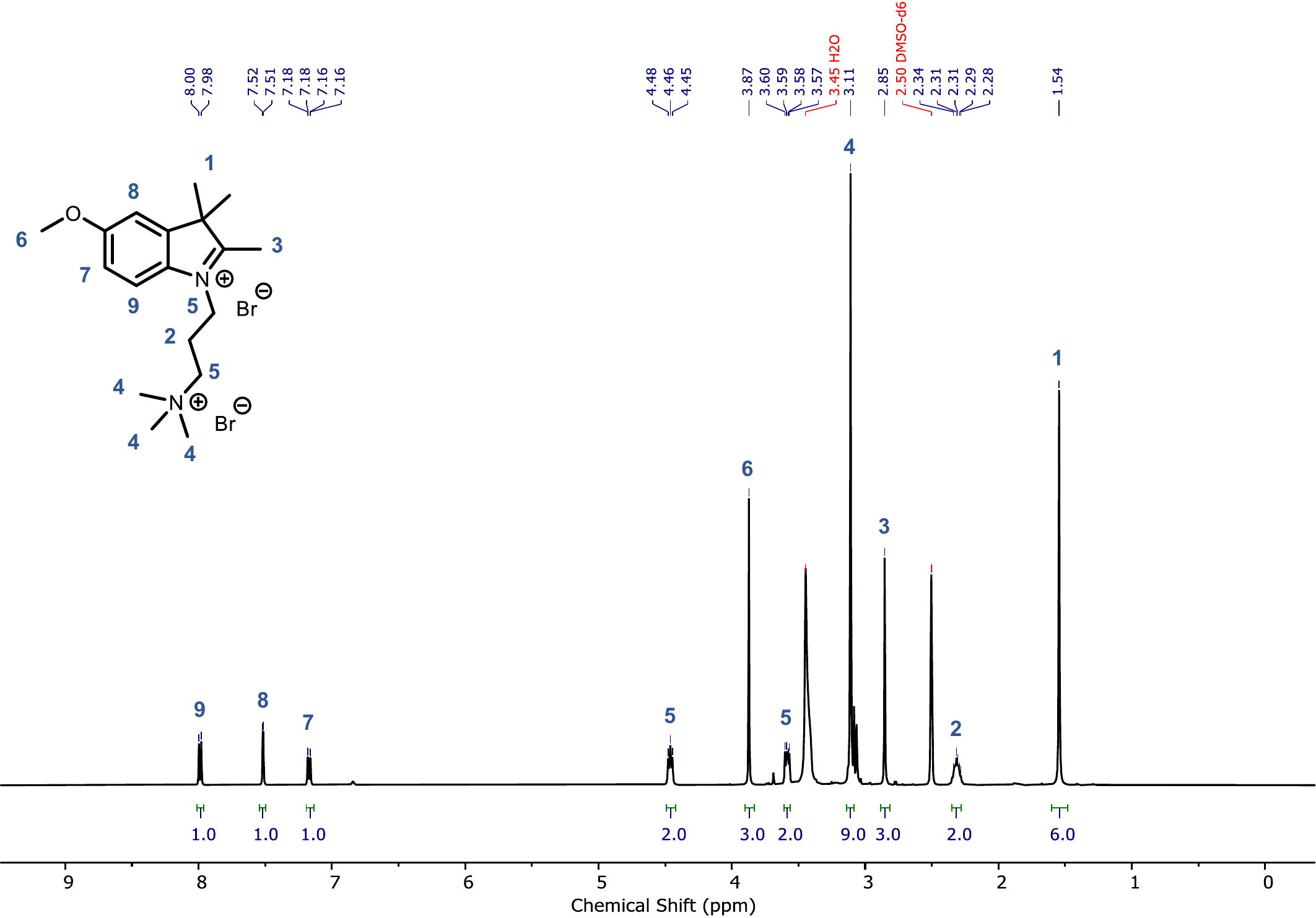}
Figure S3: {\textsuperscript{1}H NMR spectrum of 5-Methoxy-2,3,3-trimethyl-1-(3-(trimethylammonio)propyl)-3H-indol-1-ium dibromide.}
\label{Fig. S3}
\end{figure*}

\subparagraph{(E)-2-(2-hydroxy-4-methoxystyryl)-5-methoxy-3,3-dimethyl-1-(3-(trimethylammonio)propyl)-3H-indol-1-ium dibromide) (MCH-para)}\mbox{}\\
\begin{figure*}[h]
\centering
\includegraphics[width=10.5cm]{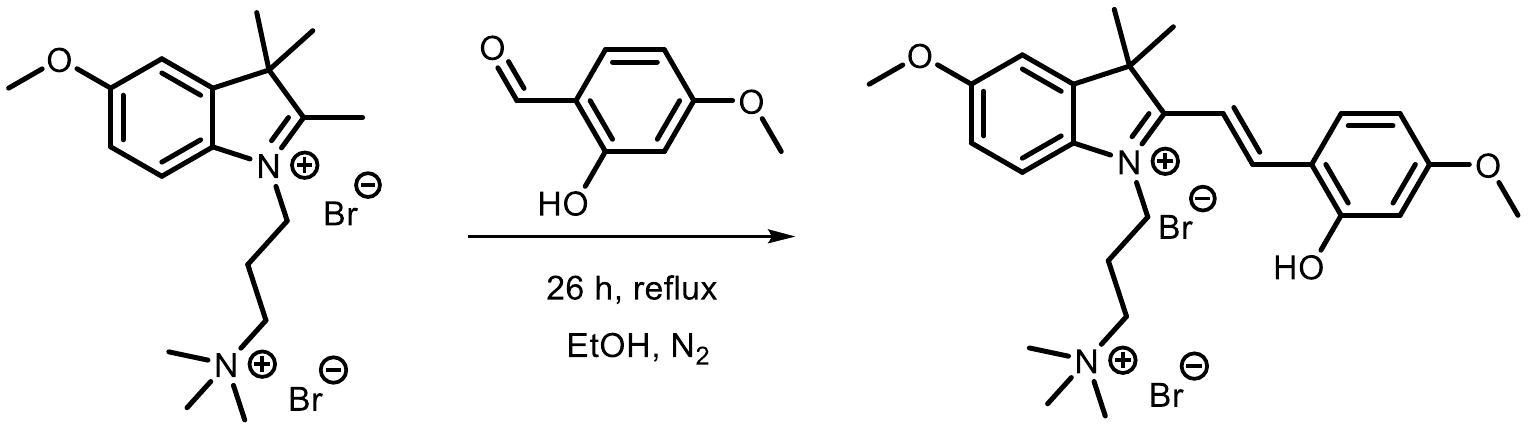}
\end{figure*}

To an oven-dry three-neck round-bottom flask equipped with a magnetic stir bar and condenser was added 2-hydroxy-4-methoxybenzaldehyde (0.30 g, 1.96 mmol, 1.4 eq.) to a solution of 5-methoxy-2,3,3-trimethyl-1-(3-(trimethylammonio)propyl)-3H-indol-1-ium dibromide (0.7 g, 90 wt\%, 1.40 mmol, 1.00 eq.), in 10 mL of anhydrous ethanol under nitrogen atmosphere. The reaction mixture was heated to 100 $^{\circ}$C and refluxed under an inert atmosphere for 26 hours. After letting the solution cool down to room temperature, it was precipitated in cold diethyl ether and washed with a considerable amount of cold dichloromethane or acetone to remove the excess amount of 2-hydroxy-4-methoxybenzaldehyde. 0.52 g (68\% yield).
\subparagraph{Merocyanine form:}\mbox{} \\
\textbf{\textsuperscript{1}H NMR} (500 MHz, DMSO-\emph{d\textsubscript{6}}:) $\delta$ 11.25 (s, 1H), 8.49 (d, \emph{J} = 16.0 Hz, 1H), 8.20 (d, \emph{J} = 8.9 Hz, 1H), 7.87 (d, \emph{J} = 8.9 Hz, 1H), 7.53 (d, \emph{J} = 2.5 Hz, 1H), 7.47 (d, \emph{J} = 16.1 Hz, 1H), 7.16 (dd, \emph{J} = 8.8, 2.5 Hz, 1H), 6.64 (dd, \emph{J} = 8.9, 2.5 Hz, 1H), 6.60 (d, \emph{J} = 2.4 Hz, 1H), 4.57 (t, \emph{J} = 7.8 Hz, 2H), 3.89 (s, 3H), 3.85 (s, 3H), 3.64 – 3.58 (m, 2H), 3.11 (s, 9H), 2.28 (td, \emph{J} = 10.1, 6.0 Hz, 2H), 1.76 (s, 6H) ppm. \\
\textbf{\textsuperscript{13}C NMR} (126 MHz, DMSO-\emph{d\textsubscript{6}}:) $\delta$ 179.63, 165.90, 161.45, 160.31, 147.71, 145.10, 134.07, 132.04, 115.65, 115.17, 114.65, 108.94, 108.23, 108.13, 100.67, 62.06, 56.20, 55.76, 52.55, 51.56, 43.11, 26.83, 21.73 ppm.\\

\begin{figure*}[!htb]
\centering
\includegraphics[width=15cm]{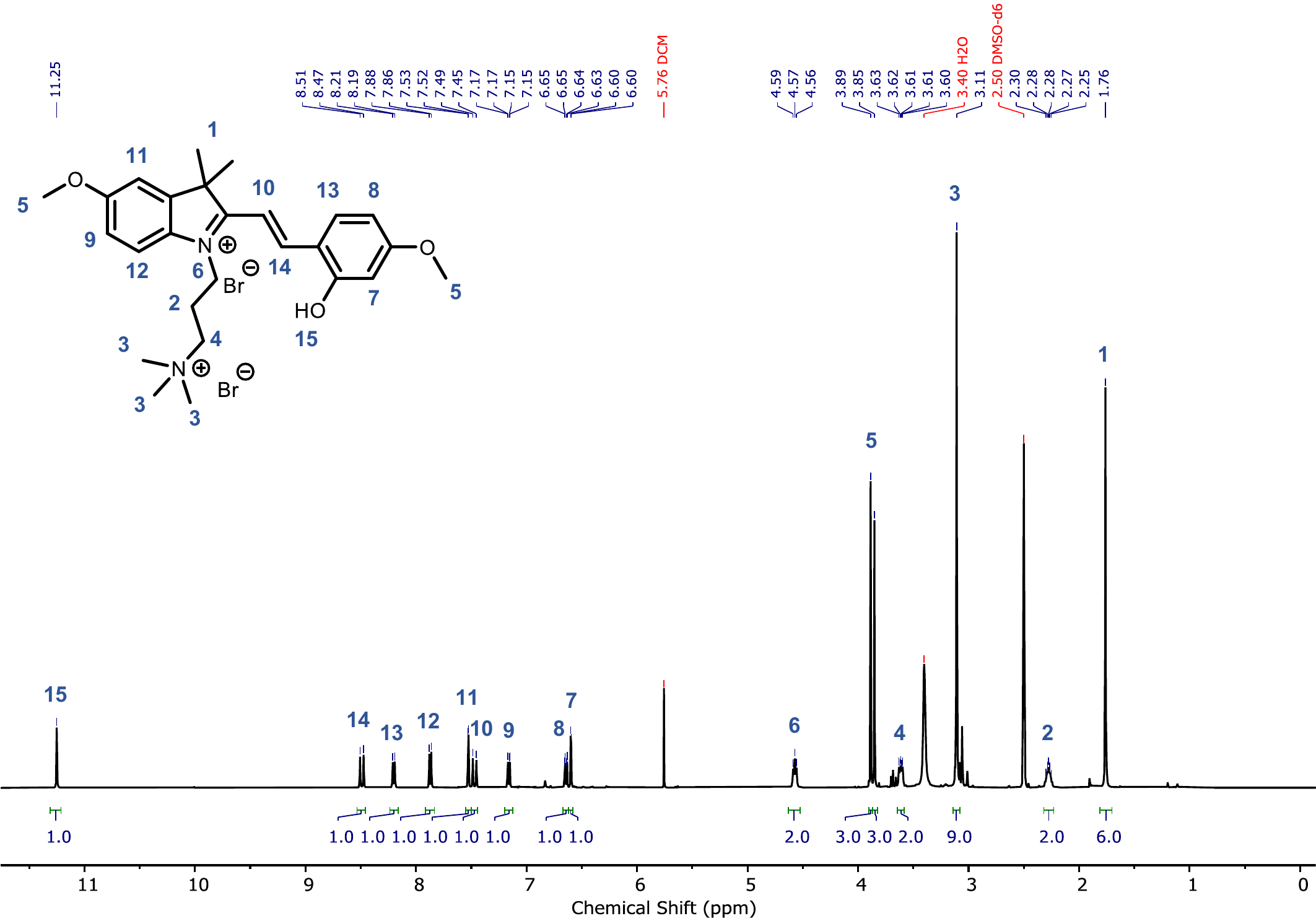}
Figure S4:{\textsuperscript{1}H NMR spectrum of 5-Methoxy-2,3,3-trimethyl-1-(3-(trimethylammonio)propyl)-3H-indol-1-ium dibromide.}
\label{Fig. S4}
\end{figure*}

\begin{figure*}[!htb]
\centering
\includegraphics[width=15cm]{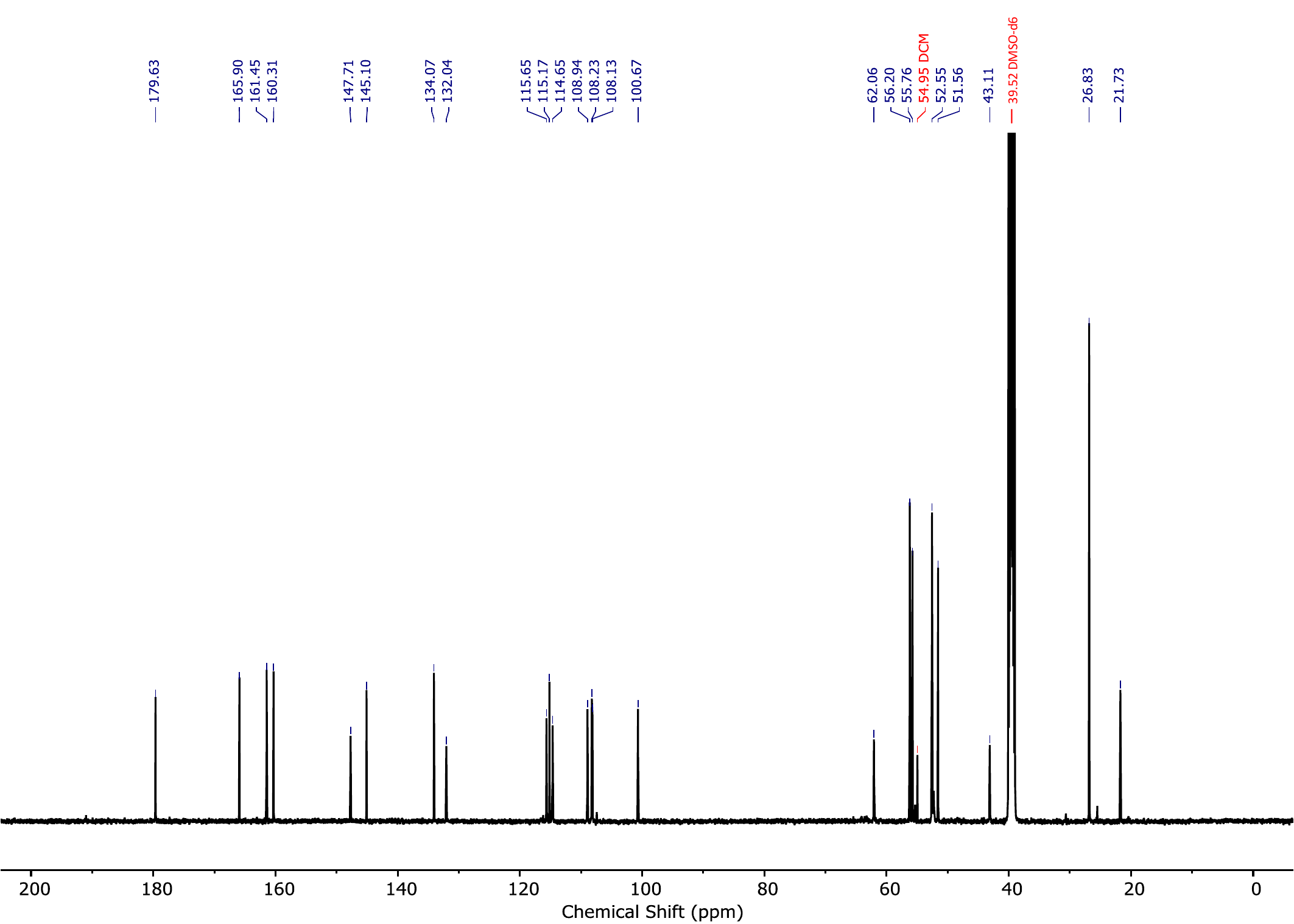}
Figure S5: {\textsuperscript{1}H NMR spectrum of 5-Methoxy-2,3,3-trimethyl-1-(3-(trimethylammonio)propyl)-3H-indol-1-ium dibromide.}
\label{Fig. S5}
\end{figure*}

\clearpage
\subsection*{Stability of SP-DA-PEG and MCH-para in water solutions}\mbox{} \\
The stability of SP-DA-PEG and MCH-para is estimated by recording UV–Vis absorption spectra with an Agilent 8453 UV–Vis spectrometer and measuring surface tension with a commercial tensiometer (Theta Flex, Biolin Scientific).
The 0.05 mM SP-DA-PEG solution was prepared as outlined in the methods and stored in a cool, dark environment. Weekly measurements of its UV–Vis spectra are depicted in SI Appendix, Figure S6 A. After eight weeks, the SP peak shifts and decreases, indicating hydrolysis-induced degradation. Additionally, using the standard pendant drop method, the change in surface tension of a fresh SP-DA-PEG aqueous solution was measured in an inert gas, nitrogen. As shown in Figure S6 B, the peak decreases as the exposure time increases, indicating that excessive exposure can hasten degradation.  
\\
\begin{figure*}[h]
\centering
\includegraphics[width=\linewidth]{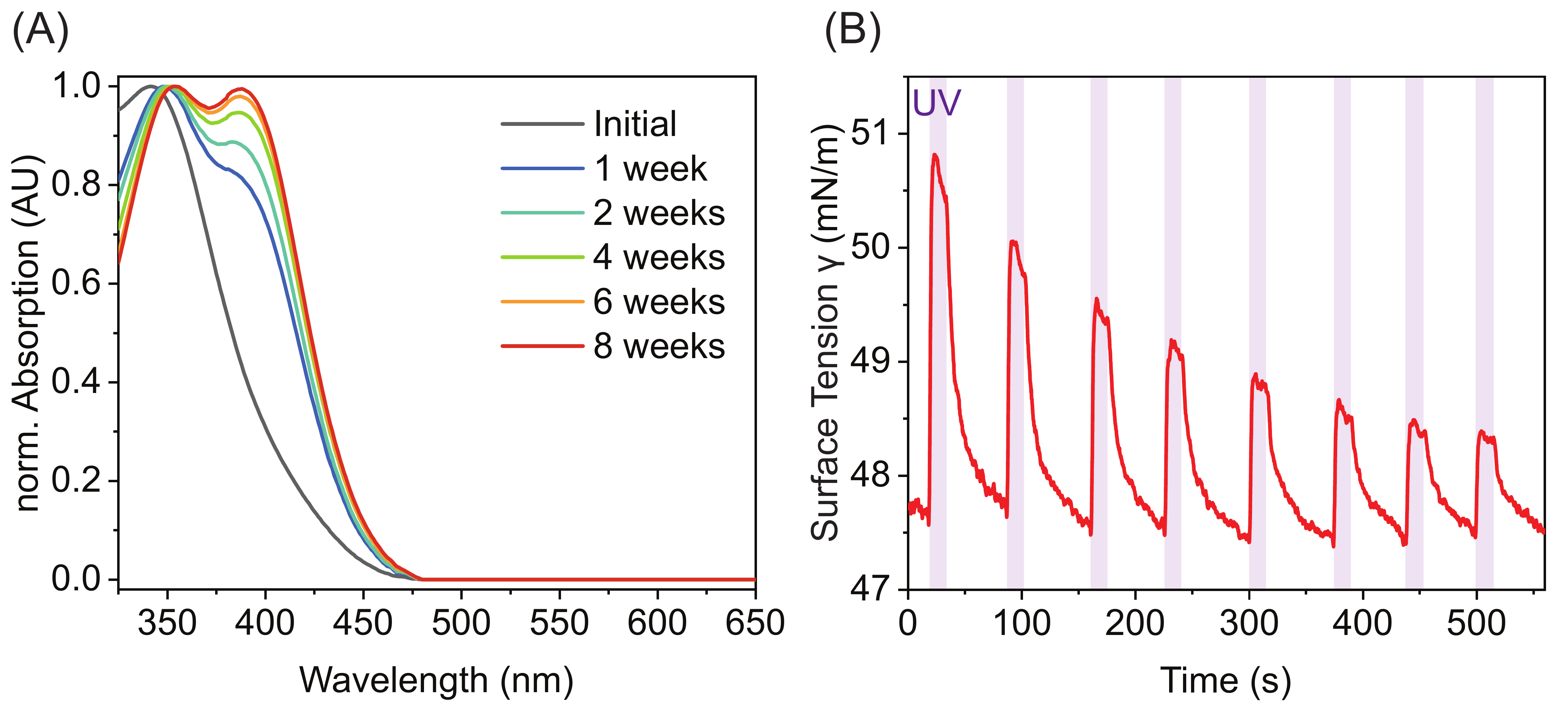}
Figure S6: {(A) UV-vis absorbance of SP-DA-PEG in water at 0.05 mM over eight weeks. (B) Surface tension of a solution of 0.2 mM SP-DA-PEG under repeatedly pulsed UV light.}
\label{Fig. S6}
\end{figure*}\\

MCH-para is stable in acidic aqueous media at room temperature for more than one month. Spectra of UV–Vis absorbance for 1 mM solutions are recorded.  In Figure S7 A, UV-Vis absorbance spectra with a 1–10 pH range indicate that both MC and MCH states can exist in a solution depending on its acidity. The equilibrium shifts to the MCH state at pH values between 1 and 4 (Fig.~S7 B). MCH is stable at pH ranges between 1 and 3, as shown in SI Appendix, Figure S7 C, where peak decreases are all less than 10\% after six weeks.  Since pH 3 has the greatest surface tension change, for experiments involving the driving droplet motions, we utilized pH 3 solutions and evaluated the stability of the corresponding solution weekly (SI Appendix, Figure S7 D). \\

\begin{figure*}[h]
\centering
\includegraphics[width=\linewidth]{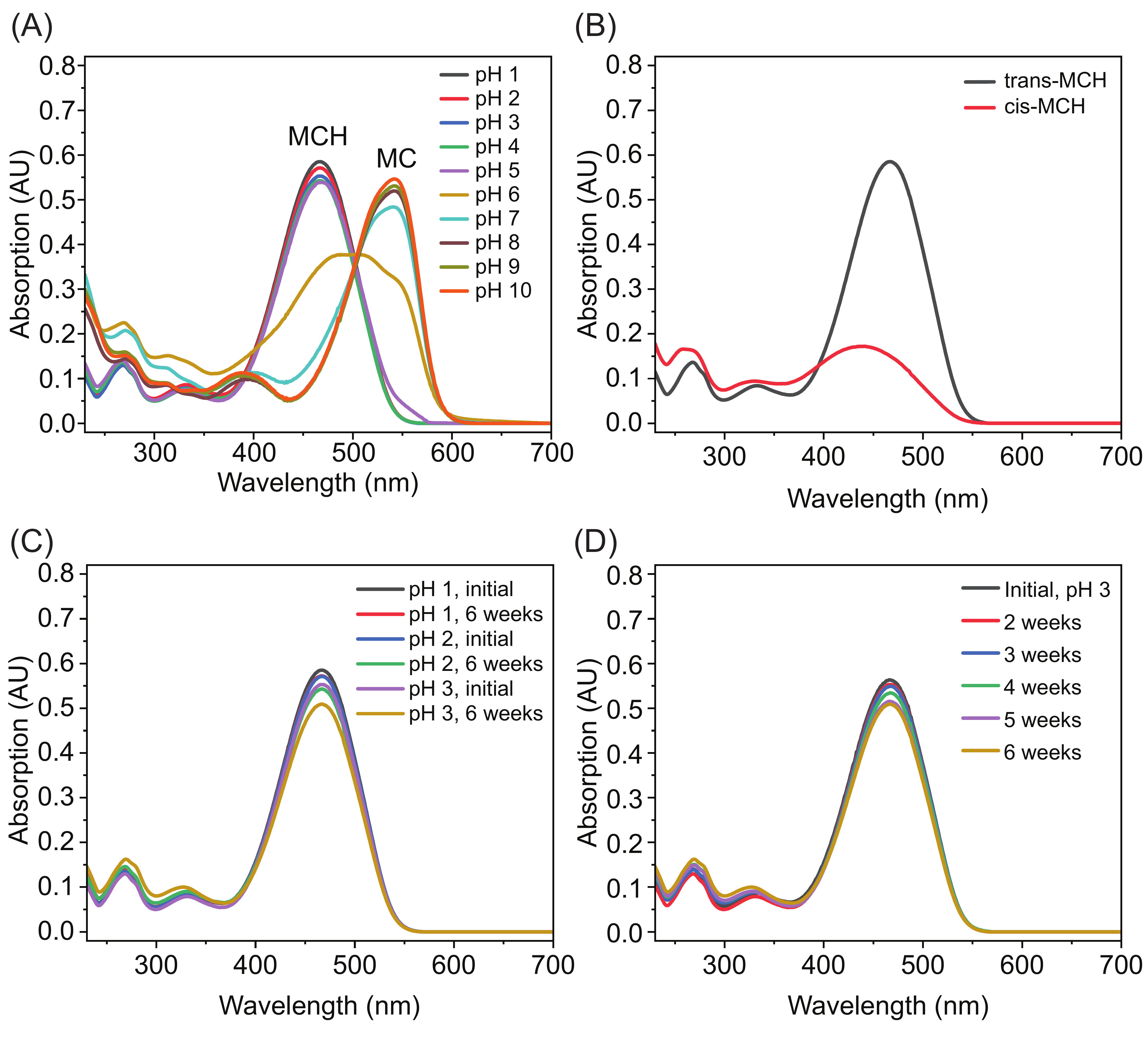}
Figure S7: {(A) UV-Vis absorbance spectra of equilibrated coexisting states of MCH-para aqueous solution over the pH range of 1 to 10. MCH form is the predominant form in low pH solutions, whereas MC is the predominant form in high pH solutions. (B) Switching mechanism of MCH-para in aqueous solution (pH 1). (C) UV-Vis spectra of trans-MCH para in low pHs. Red line and blue line overlap. (D) UV-Vis spectra of MCH-para in aqueous solution at pH 3 from week 1 to week 6.}
\label{Fig. S7}
\end{figure*}

\clearpage
\subsection*{Temperature rise of droplet due to heating from light illumination}\mbox{} \\
It is possible for the thermal Marangoni effect induced by irradiating the substances with light to partially contributes to the droplet motion. Surface tension of an aqueous solution containing the photo-responsive surfactants as a function of temperature was measured using the standard pendant drop method on a commercial tensiometer. The increase in temperature caused by illumination of light with specific intensity was estimated using an infrared (IR) camera (Telops M3K). 
To decouple the photo Marangoni effect from heating effect under illumination, we first heated the solution to varying temperatures using a custom syringe heating system and then measured the surface tension. Several film heater plates were attached to the exterior of the syringe to heat the solution, and a thermocouple was attached to the needle to measure the temperature of the injected droplet (SI Appendix, Figure S8 A). As a result, the relationship between surface tension and temperature is linear, ranging from 20~$^{\circ}$C to 60 $^{\circ}$C (SI Appendix, Figure S8 B). As temperature increases, surface tension decreases: \\
\begin{equation}
 \gamma \left[\frac{\text{mN}}{\text{m}} \right] = -0.26\; T\,[^\circ \text{C}] + 76.66 
\label{eqn1}
\end{equation}
\\
To further assess the thermal effect, a series of temperature measurement experiments have been conducted. The emissivity of an aqueous solution containing 1 mM MCH-para was calibrated with reference to the emissivity of deionized water.  Images of the two droplets show comparable temperature profiles and heat distributions when two identically sized droplets of deionized water and MCH-para solution are subjected to the same room temperature and light conditions (SI Appendix, Figure S8 C). As a result, the emissivity of water, $\epsilon$ = 1, can be used to estimate the emissivity of MCH-para. 
The temperature rise in MCH-para droplet due to blue LED illumination with an intensity of 31.8~mW/cm\textsuperscript{2} is small. The collimated 470 nm LED used to illuminate the droplet is the same one utilized for measuring surface tension (SI Appendix, Figure S9 A). As comparison, the LED was first used to heat a water droplet of equal size, resulting in a temperature change of less than 1~$^{\circ}$C (SI Appendix, Figure S9 B). Then, the MCH-para droplet was stabilized at room temperature. After 150~s of heating, the maximum temperature within the droplet rises from 19.6 $^{\circ}$C to 20.6 $^{\circ}$C (SI Appendix, Figure S9 C and D), resulting in a 0.26~mN/m decrease in surface tension. This decrease is negligible compared to the total change of 4.4 mN/m in surface tension (pH = 3) depicted in Figure 1H. The thermal effect caused by irradiation only accounts for 5.9\% of the total change in surface tension. For blue light with similar intensities, the photo Marangoni effect is the primary factor that induces shear flow and causes droplet motions. 

\section*{Simulations of drops over liquid-infused surfaces}

In order to help understand the flow patterns arising within drops on LIS, we performed simulations using COMSOL Multiphysics. The  simulations  solved  the Navier-Stokes equations for the transport of mass and momentum in the fluid. For simplicity, the aqueous drop shape is approximated as a hemisphere, and is placed on an oil layer of uniform thickness. The flow is assumed steady in a reference frame moving at a velocity $V$ with drop. Therefore, at the oil-water interface we enforce no normal flow, and we enforce uniform flow, equal to $-V$,  in the oil layer sufficiently far away from the drop. The bottom of the oil layer has a no-slip boundary condition in the fixed reference frame, and therefore has a velocity equal to $-V$  when observed in the frame of reference moving with the drop. We enforce continuity of tangential velocity at the drop-oil interface. The effect of the Marangoni stress is captured by applying a meridional stress along the bottom and top boundaries of the drop. A refined mesh at the interface and near the endpoints is used to ensure numerical accuracy.

\newcommand{\Ca}{C_\text{A}}
\newcommand{\Cb}{C_\text{B}}
\newcommand{\CaI}{\left.C_\text{A}\right|_\text{I}}
\newcommand{\CbI}{\left.C_\text{B}\right|_\text{I}}
\newcommand{\aAB}{a_\text{AB}}
\newcommand{\aBA}{a_\text{BA}}
\newcommand{\kAbsA}{\kappa_\text{ads}^\text{A}}
\newcommand{\kAbsB}{\kappa_\text{ads}^\text{B}}
\newcommand{\kDesA}{\kappa_\text{des}^\text{A}}
\newcommand{\kDesB}{\kappa_\text{des}^\text{B}}
\newcommand{\gradI}{\vec{\nabla}_\text{I}}
\newcommand{\uI}{\vec{u}_\text{I}}
\newcommand{\vel}{\vec{u}}
\newcommand{\Iba}{I_\text{BA}}
\newcommand{\Iab}{I_\text{AB}}
\newcommand{\ud}{\text{d}}

The pure photo-switch is
\begin{equation}
    \frac{\partial{\Ca}}{\partial{t}}=-\aBA\Ca+\aAB{\Cb}=-\frac{\partial{\Cb}}{\partial{t}}
\end{equation}
where 
\begin{align}
    \aBA &= \frac{\ud \aBA}{\ud\Iba} \,\Iba\\
    \aAB &= \frac{\partial \aAB}{\partial \Iab} \, \Iab + \aAB^\text{dark}
\end{align}
Bulk:
\begin{align}
\vec{\nabla}\cdot\vel &=0, \\
    \frac{\partial{\vel}}{\partial{t}}+\vec{\nabla}\cdot(\vel \,\vel)&=\nu\nabla^2\vel-\rho^{-1}\vec{\nabla}p, \\
    \frac{\partial{\Ca}}{\partial{t}}+\vec{\nabla}\cdot(\vel\,\Ca)&=D\nabla^2\Ca-\aAB{\Ca}+\aBA{\Cb}, \\
    \frac{\partial{\Cb}}{\partial{t}}+\vec{\nabla}\cdot(\vel\,\Cb)&=D\nabla^2\Cb+\aAB{\Ca}-\aBA{\Cb}.
\end{align}
On interface:
\begin{align}
  \gamma & =\gamma_o+n R T\Gamma_\text{m}\ln \left(1-\frac{\Gamma_\text{A}+\Gamma_\text{B}}{\Gamma_\text{m}} \right)\\
\frac{\partial{\Gamma_\text{A}}}{\partial{t}}+\gradI \cdot(\uI 
 \Gamma_\text{A}) & =D\nabla_\text{I}^2\Gamma_\text{A}+S_\text{A}-\aAB{\Ca}+\aBA{\Cb}\\
\frac{\partial{\Gamma_\text{B}}}{\partial{t}}+\gradI\cdot(\uI \Gamma_\text{B}) & =D\nabla_\text{I}^2\Gamma_\text{B}+S_\text{B}+\aAB{\Ca}-\aBA{\Cb}\\
S_\text{A} & =\kAbsA \, \CaI [\Gamma_\text{m}-(\Gamma_\text{A}+\Gamma_\text{B})]- \kDesA \,\Gamma_\text{A}\\
S_\text{B} & =\kAbsB \, \CbI [\Gamma_\text{m}-(\Gamma_\text{A}+\Gamma_\text{B})]-\kDesB \,\Gamma_\text{B}\\
-D\left.\frac{\partial{\Ca}}{\partial{n}}\right|_\text{I}&=S_\text{A}\\
-D\left.\frac{\partial{\Cb}}{\partial{n}}\right|_\text{I}&=S_\text{B}\\
\gradI \gamma &= \mu \left[ \left.\frac{\partial \vec{u}}{\partial{n}}\right|_{\text{I}+} - \left.\frac{\partial \vec{u}}{\partial{n}}\right|_{\text{I}-} \right]
\end{align}

\section*{Scaling theory for drops in liquid}
\newcommand{\eul}{\text{e}}
\newcommand{\tFwd}{\tau_\text{fwd}}
\newcommand{\tRev}{\tau_\text{rev}}
\newcommand{\tIll}{t_\text{i}}
\newcommand{\tDark}{t_\text{d}}
\newcommand{\gammaA}{\gamma_\text{A}}
\newcommand{\gammaB}{\gamma_\text{B}}
\newcommand{\gammaLo}{\gamma_\text{lo}}
\newcommand{\gammaHi}{\gamma_\text{hi}}
\newcommand{\gammaOne}{\gamma_\text{1}}
\newcommand{\VI}{V_\text{I}}
The Marangoni stress across the drop is of order $(\gammaHi-\gammaLo)/R$, where $\gammaHi$ and $\gammaLo$ are the interfacial tension values at the front and rear of the drop, whereas $R$ is the drop radius. The drop surface area is of order $R^2$; therefore
\begin{equation}
    F_\text{Ma} \sim (\gammaHi-\gammaLo) R.
\end{equation}

At low Reynolds numbers $Re = \rho V 2R/\mu$ (where $\rho,\mu$  are the Krytox density and viscosity), the fluid resistance associated with a self-propelled drop in a fluid whose viscosity is much greater than the drop's is of order $F_\text{drag} \sim \mu V R$ \cite{Bjelobrk2016-sb}.  Therefore the balance $F_\text{drag} \sim F_\text{Ma}$ yields
\begin{equation}
    V \sim \frac{\gammaHi-\gammaLo}{\mu}\quad\text{if}\quad\mu\gg\mu_\text{drop}, Re\ll1.
    \label{eq:VloReStart}
\end{equation}
We must determine how the difference $\gammaHi-\gammaLo$ depends on other parameters, such as drop radius.  Let us label $\gammaA$ the tension in the rest (dark) state, whereas $\gammaB$ is the steady-state value eventually reached under illumination. Empirically, based on our surface tension measurements, we approximate the forward switch dynamics (starting in the dark and illuminating at $t=0$)
\begin{equation}
\gamma(t) = \gammaB + (\gammaA-\gammaB)\, \eul^{-t/\tFwd},
\end{equation}
where $\tFwd$ is the timescale for the forward switch. Similarly, when illumination is stopped from the initial steady state $\gamma = \gammaB$, the interfacial tension follows approximately
\begin{equation}
    \gamma(t) = \gammaA + (\gammaB-\gammaA)\, \eul^{-t/\tRev}
    \label{eq:gammaBtoA}
\end{equation}
where $\tRev$ is the timescale for the reverse switch. Here, $\gammaA, \gammaB, \tRev$ and $\tFwd$ are found empirically from surface-tension measurements, such as those in Fig.~\ref{Fig. S6}B.

To find expressions for $\gammaHi$ and $\gammaLo$, we derive a set of equations relating the interfacial tensions at key locations on the drop. We label the solution the leaves the illuminated region of the drop as $\gammaHi$. This liquid takes a time $\tDark \sim (2R-d)/\VI$  to reach the rear end of the drop, where $\VI$ is the fluid velocity scale at the interface and in the interior of the drop. Therefore, noting that the initial condition is $\gamma = \gammaHi$, Eq.~\ref{eq:gammaBtoA} becomes:
\begin{equation}
    \gammaLo \sim \gammaA + (\gammaHi-\gammaA)\,\eul^{-\tDark/\tRev}
\label{eq:gammaLo}
\end{equation}
where $d$ is the light beam diameter. As the liquid recirculates inside the drop, the photo-surfactant continues its reverse reaction. Even though this liquid is not at an interface, we can model the value of the interfacial tension that would arise if this fluid element were placed at an interface (for brevity, we label this a `virtual tension'). Since the liquid has traveled over a time of order $2\tDark$ since it was last illuminated, when it is about to enter the illuminated region again it will have a virtual tension, labeled here $\gammaOne$,
\begin{equation}
    \gammaOne \sim \gammaA + (\gammaHi-\gammaA)\, \eul^{-2 \tDark/\tRev}
    \label{eq:gammaOne}
\end{equation}
The highest surface tension is reached by illumination of this fluid as it travels a length of order $2d$, over a time $\tIll \sim 2d/(VI\tRev)$, 
\begin{equation}
\gammaHi \sim \gammaB + (\gammaOne-\gammaB)\,\eul^{-\tIll/\tRev}
\label{eq:gammaHi}
\end{equation}
We use Eqs.~\eqref{eq:gammaLo}, \eqref{eq:gammaOne} and \eqref{eq:gammaHi} in Eq.~\eqref{eq:VloReStart} and eliminate $\gammaOne, \gammaHi$, and $\gammaLo$, obtaining:
\begin{equation}
V_{Re\ll1} \sim \frac{\gammaHi - \gammaLo}{\mu} \sim \frac{\gammaB-\gammaA}{\mu}\,\dfrac{(1-\eul^{-\tDark/\tRev})(1-\eul^{-\tIll/\tFwd})}{1-\eul^{-(2{\tDark}/{\tRev}+{\tIll}/{\tFwd})}}.
\label{eq:VloReWithVI}
\end{equation}
If $Re\ll1$, we can assume $V_I\sim V$. Introducing  scaling constants $c_1$ and  $c_2$, we obtain an implicit equation for $V$:
\begin{equation}
V_{Re\ll1} = c_1 \frac{\gammaB-\gammaA}{\mu}\,\dfrac{(1-\eul^{-c_2\tDark/\tRev})(1-\eul^{-c_2\tIll/\tFwd})}{1-\eul^{-c_2(2{\tDark}/{\tRev}+{\tIll}/{\tFwd})}}. 
\label{eq:VloRe}
\end{equation}
Eq.~\eqref{eq:VloRe} above is solved by iteration and is used to plot the dotted line in Fig.~4E. We use $\gammaB-\gammaA = 0.73$\,mN/m, as this is the value observed after repeated illumination cycles (as exemplified in Fig.~\ref{Fig. S6}B), and $\tFwd = 0.437\,\text{s}, \tRev = 2.19\,\text{s}, \mu = 1.24\cdot 10^{-2}$/\,kg/(m\,s), $d=1.3$\,mm and $R = (\frac{4}{3\pi}\mathcal{V})^{1/3}$, where $\mathcal{V}$ is the drop volume. The constants we use are  $c_1 = 0.2$ and $c_2=0.15$; we note these are of order-one, as should be expected for a physically plausible scaling theory.

It is instructive to seek an approximation of Eq.~\eqref{eq:VloRe} for small drops. We assume that, as $R$ approaches zero, $V$ decreases more slowly that $R$, such that the ratio $R/(VI\,\tRev)$ also approaches zero. Since in practice the beam diameter $d$ is essentially constant, we assume that the ratio $d/(VI\,\tFwd) \sim \tIll/\tRev \gg 1$, such that, in Eq~\eqref{eq:VloRe}, we neglect the second exponential in the numerator, as well as the exponential in the denominator:
\begin{equation}
    V_{Re\ll1} \approx c_1 \frac{\gammaB-\gammaA}{\mu}\,(1-\eul^{-c_2\tDark/\tRev}).
\end{equation}
Introducing a Taylor expansion in $\tDark/\tRev$, neglecting second-order terms, and solving for $V$, we obtain the explicit expression
\begin{equation}
    V_{Re\ll1} \approx  \sqrt{c_1 c_2 \frac{\gammaB-\gammaA}{\mu}\, \dfrac{2R-d}{\tRev}},
    \label{eq:VscaleSmallDrop}
\end{equation}
suggesting that for small drops the velocity is highly sensitive to volume, as $V \sim R^{1/2}\sim\mathcal{V}^{1/6}$. 

At large drop volumes, we observe empirically that the velocity decreases slightly. There are several potential explanations for this observation. We tentatively consider here the effect of finite Reynolds number, as the larger drops have $Re = \rho V 2R/\mu \sim 4$. For a self-propelled object, the propulsive thrust $T$ scales as $T \sim \dot{m} \Delta V$, where $\dot{m}$ is the mass flow rate that is accelerated by a velocity increment $\Delta V$. Here $\dot{m} \sim \rho (V+\Delta V) \delta^2$, where $\delta$ is the half-width of the faster-moving fluid region behind the drop. Equating the thrust with the Marangoni force $\sim (\gammaHi-\gammaLo)R$ we find
\begin{equation}
    (\gammaHi-\gammaLo)R \sim \rho (V+\Delta V) \Delta V \delta^2.
\end{equation}
For large $Re\gg1$, we use the canonical laminar flow scaling $\delta^2 \sim R^2 \, Re_{\Delta V}^{-1} \sim R \nu \Delta V$. Assuming $\Delta V \sim V$, the scaling simplifies to
\begin{equation}
    (\gammaHi-\gammaLo) \sim \mu \Delta V .
    \label{eq:gammaDeltaV}
\end{equation}
To relate $V$ and $\Delta V$, we assume that at $Re \gg 1$ the fluid resistance scales as $\rho V^2 R^2$. Balancing this fluid resistance with $T$,
\begin{equation}
    \rho (V+\Delta V) \Delta V \delta^2 \sim \rho V^2 R.
    \label{eq:dragScaling}
\end{equation}
Using again the assumption $\Delta V \sim V$, Eq.~\eqref{eq:dragScaling} reduces to
\begin{equation}
    \Delta V \sim V^2 \frac{R}{\nu}.
    \label{eq:DeltaV}
\end{equation}
Substituting \eqref{eq:DeltaV} into \eqref{eq:gammaDeltaV}, solving for $V$, and introducing scaling constants $C_1, C_2, C_3$
\begin{align}
       V_{Re\gg 1} &= C_1\sqrt{ \dfrac{\gammaHi-\gammaLo}{R \rho} }, \label{eq:VhighRe}\\
       \gammaHi-\gammaLo &= (\gammaB-\gammaA)\,
       \dfrac{\left[1-\exp\left(-C_2\dfrac{2R-d}{\VI\,\tRev}\right)\right]\left[1-\exp\left(-C_2 \dfrac{2d}{\VI\,\tFwd}\right)\right]}{1-\exp\left[ -C_2\left(2\dfrac{2R-d}{\VI\,\tRev}+\dfrac{2d}{\VI\,\tFwd}\right) \right]},\\
       V_I &= V_{Re\gg1}+C_3\,V_{Re\gg1}^2\dfrac{R}{\nu}.
\end{align}
The expression \eqref{eq:VhighRe} is used to plot the dashed line in Fig.~4E, with $C_1=1.2, C_2=3.9, C_3 = 0.1$. For large drops,  $V_{Re\gg 1} \sim R^{-1/2}\sim\mathcal{V}^{-1/6}$. 
A composite relation of the two Reynolds number regimes is shown by the continuous line in Fig.~4E, and is obtained using 
\begin{equation}
    V_\text{composite} =\left[ (V_{Re\ll1})^{-m} + (V_{Re\gg1})^{-m} \right]^{-1/m}
\end{equation}
where we set $m$=12.

\newpage
\begin{figure*}[h]
\centering
\includegraphics[width=\linewidth]{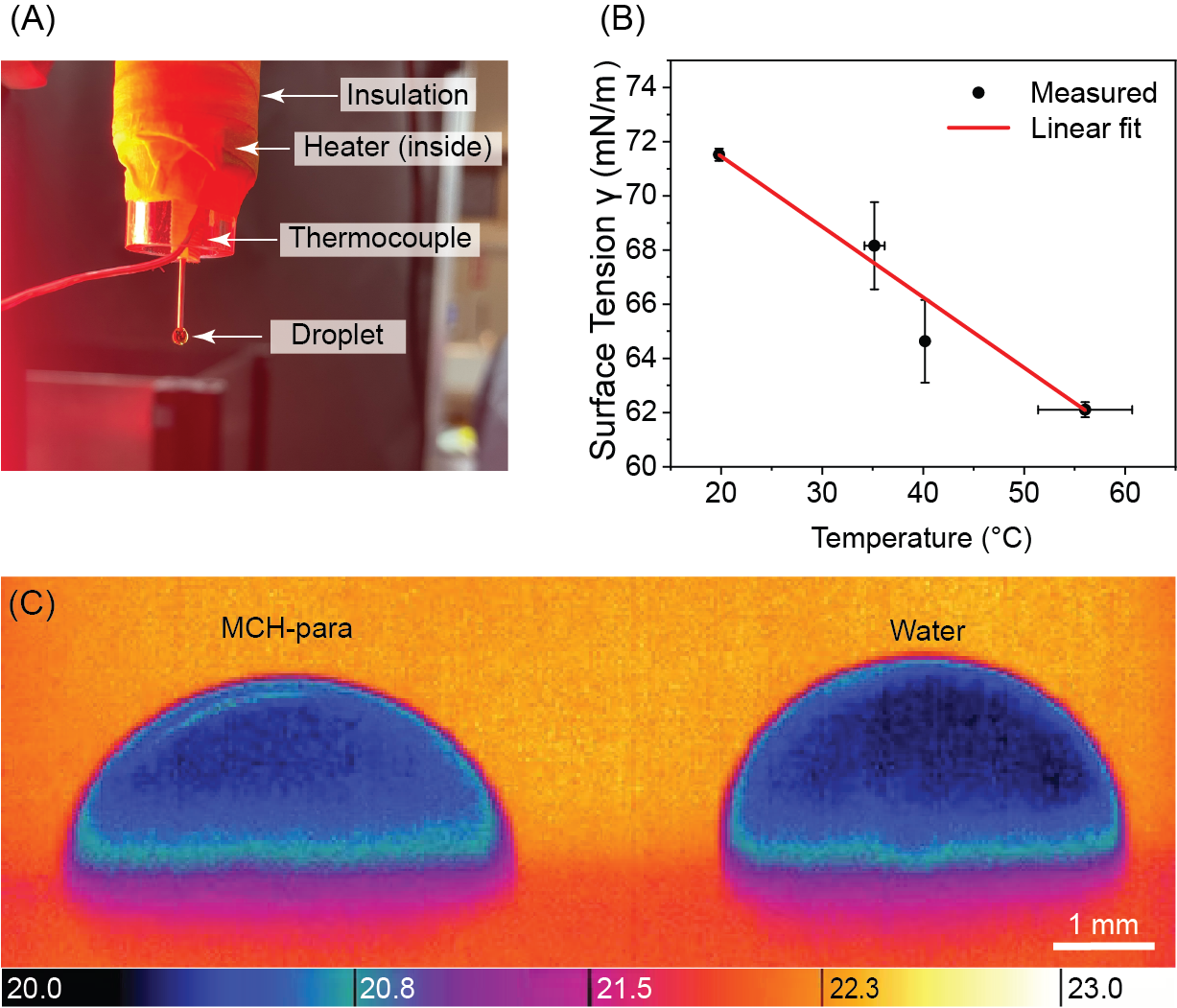}
Figure S8: {(A) Experimental setup for measuring surface tension at various temperatures. (B) The response of MCH-para surface tension to temperature. (C) IR images of the deionized water (right) and the MCH-para aqueous solution droplet (left; pH = 3) at room temperatrue. Ranges on the legend bar represent temperatures in $^{\circ}$C.}
\label{Fig. S8}
\end{figure*}

\begin{figure*}[h]
\centering
\includegraphics[width=\linewidth]{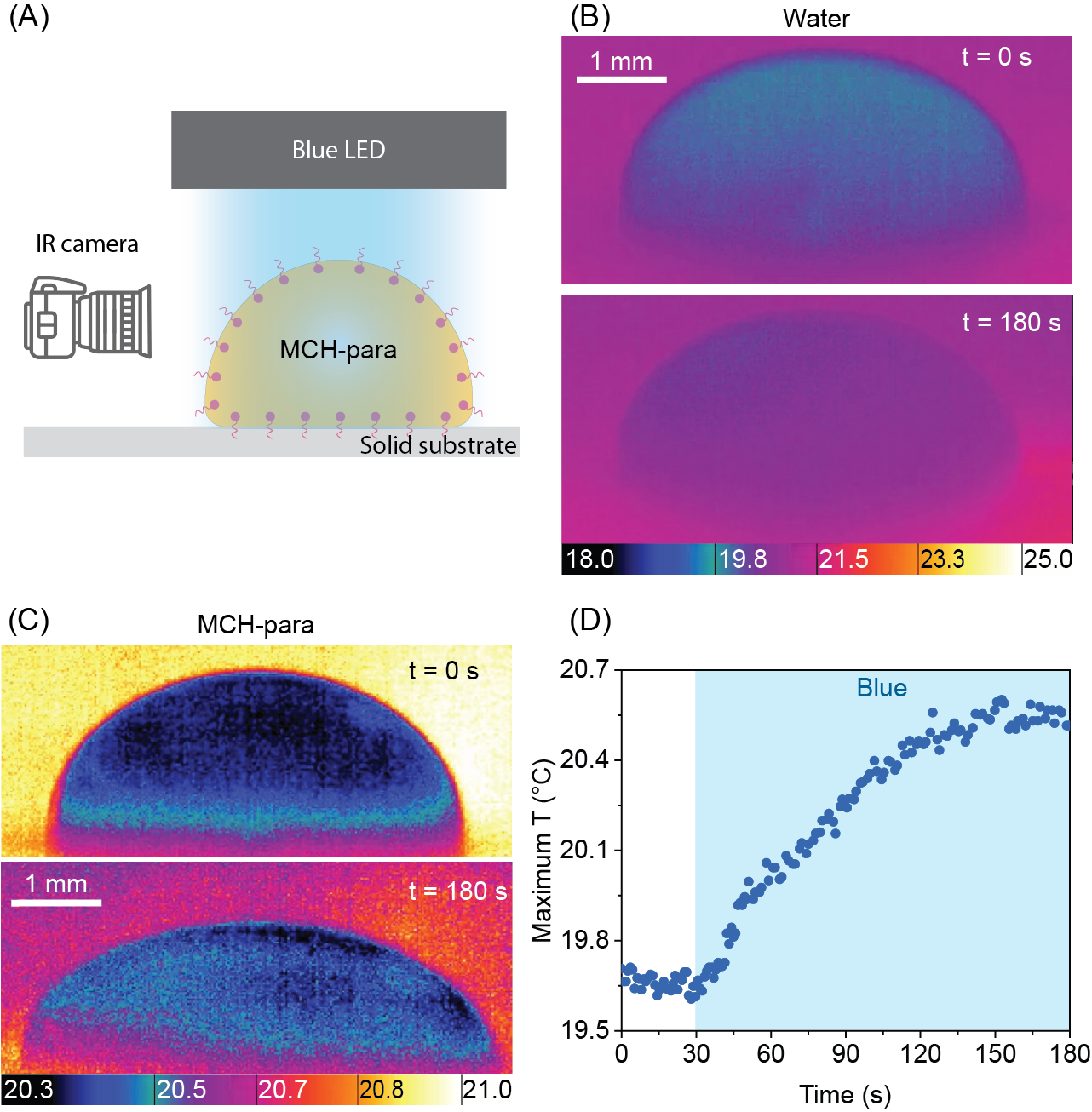}
Figure S9: {(A) Schematic of single droplet IR image capture under collimated LED illumination (31.8 mW/cm\textsuperscript{2}).  (B) IR images of blue light-heated deionized water droplets. (C) IR images of 1 mM MCH-para water droplet heated by blue light illuminations. (D) Maximum temperature occurs within the MCH-para droplet over six minutes. }
\label{Fig. S9}
\end{figure*}

\clearpage
\section*{Supplementary Figures}

\begin{figure}[!htb]
\centering
\includegraphics[width=\linewidth]{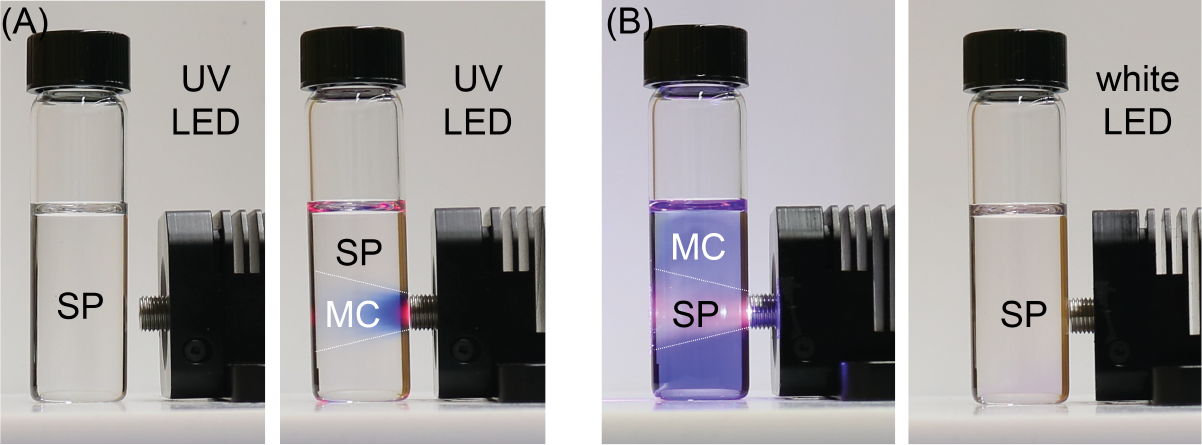}
Figure S10: {The changes in the color of SP-DA-PEG in toluene under UV illumination. (A) 0.1 mM SP toluene solution transforms from transparent (SP) to dark blue (MC) under 365 nm UV illumination. (B) After achieving equilibrium, the MC-rich solution under white light illumination changes color from dark blue to transparent. }
\label{Fig. S10}
\end{figure}

\begin{figure}[!htb]
\centering
\includegraphics[width=\linewidth]{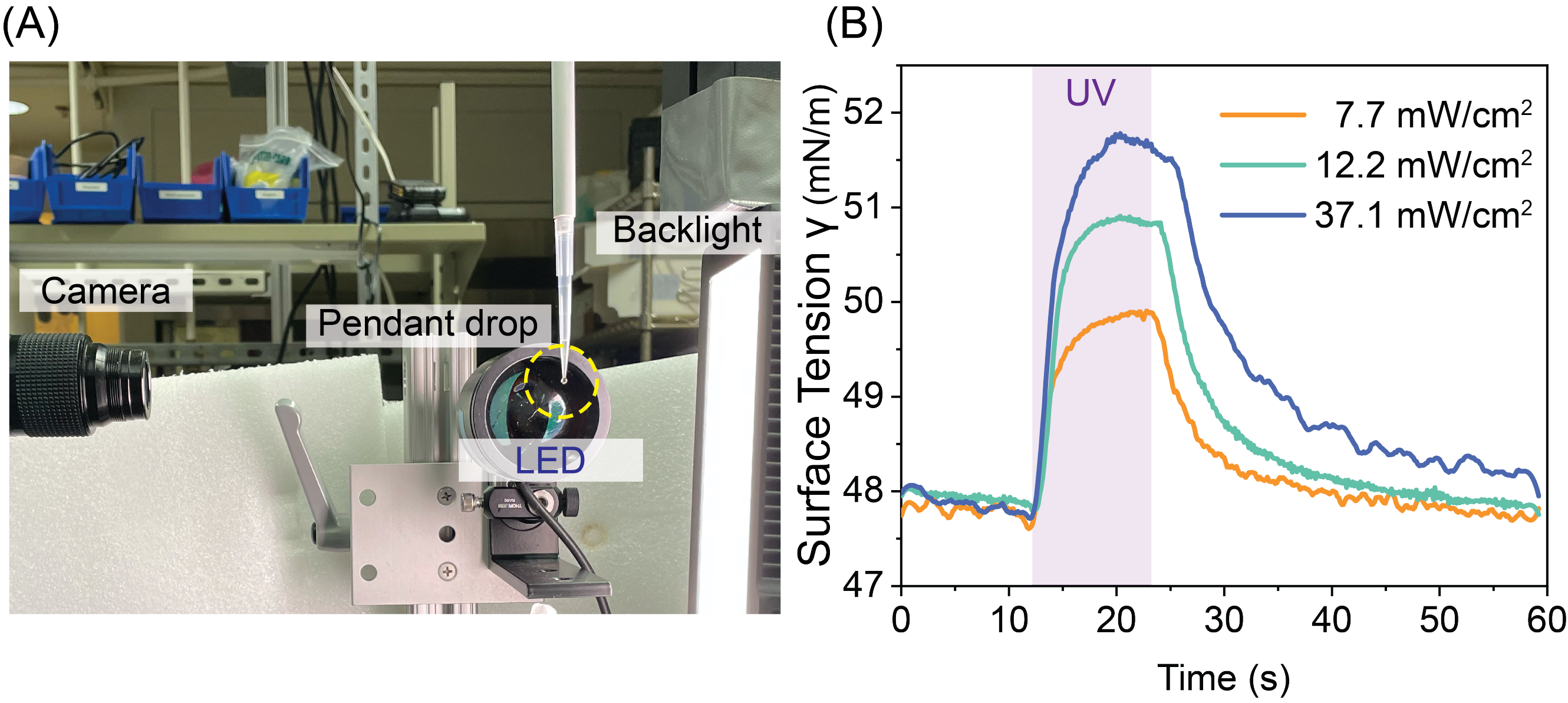}
Figure S11: {(A) Experimental setup for measuring surface tension under repeated exposure to corresponding light. (B) Surface tension response of 0.2 mM SP-DA-PEG in water under UV (365 nm) illumination with optical intensities in the range of 7.7 to 37.1 mW/cm\textsuperscript{2}.}
\label{Fig. S11}
\end{figure}

\begin{figure}[hbt!]
\centering
\includegraphics[width=\linewidth]{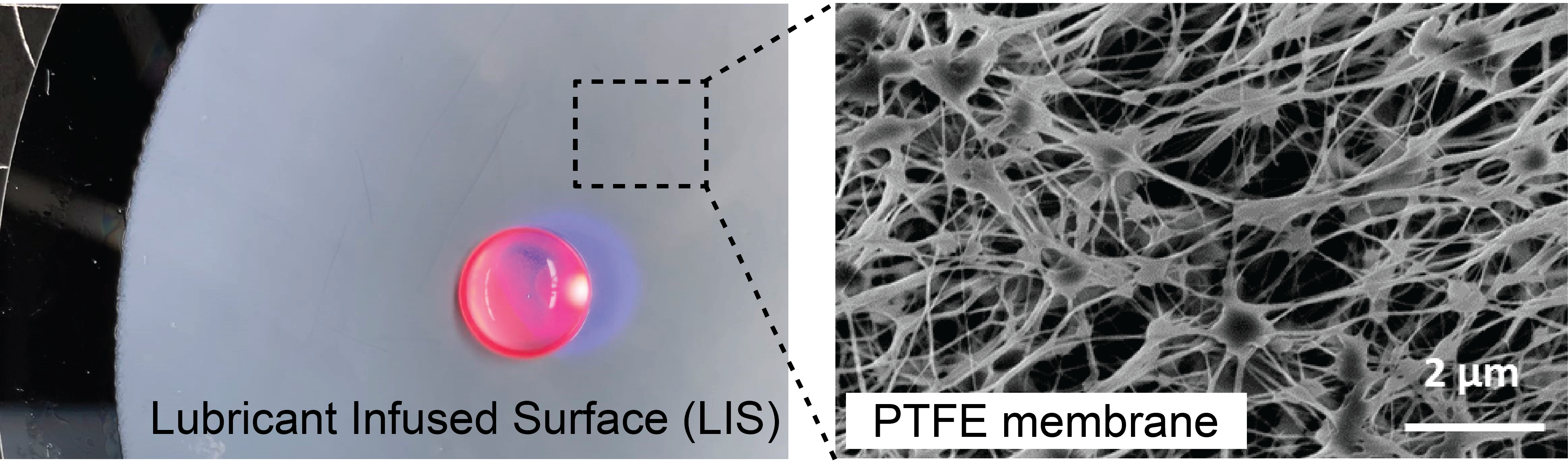}
Figure S12: {Schematic of a liquid droplet on LIS under UV illumination. Scanning Electron Microscopy (SEM) image of a porous PTFE substrate.}
\label{Fig. S12}
\end{figure}

\begin{figure}[hbt!]
\centering
\includegraphics[width=\linewidth]{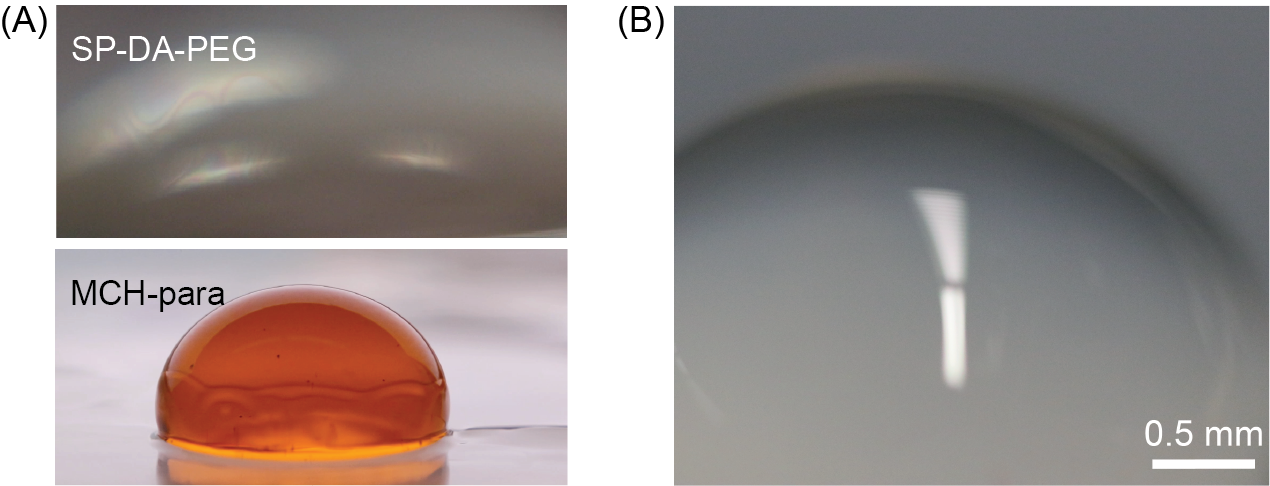}
Figure S13: {Surface of a bare 0.2 mM SP-DA-PEG water droplet placed on the PTFE film.}
\label{Fig. S13}
\end{figure}

\begin{figure}[hbt!]
\centering
\includegraphics[width=11.4 cm]{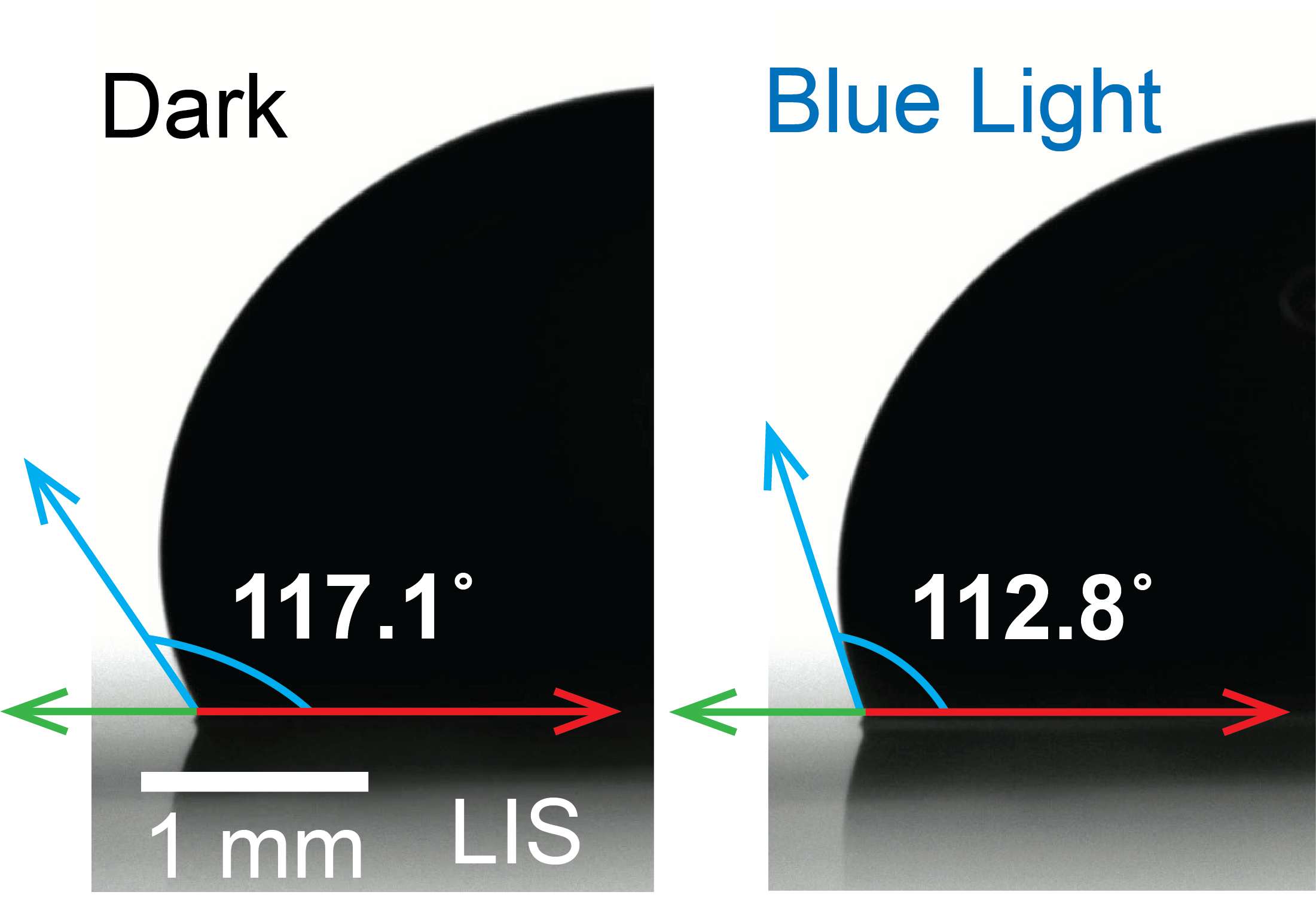}\\
Figure S14: { Contact angle measurements of 1 mM MCH-para in water at pH 3. The contact angle of MCH-para water droplet on LIS reduced from 117° to 113° upon blue light (470 nm, 31.8 mW/cm\textsuperscript{2}) irradiation.}
\label{Fig. S14}
\end{figure}

\begin{figure}[hbt!]
\centering
\includegraphics[width=11.4 cm]{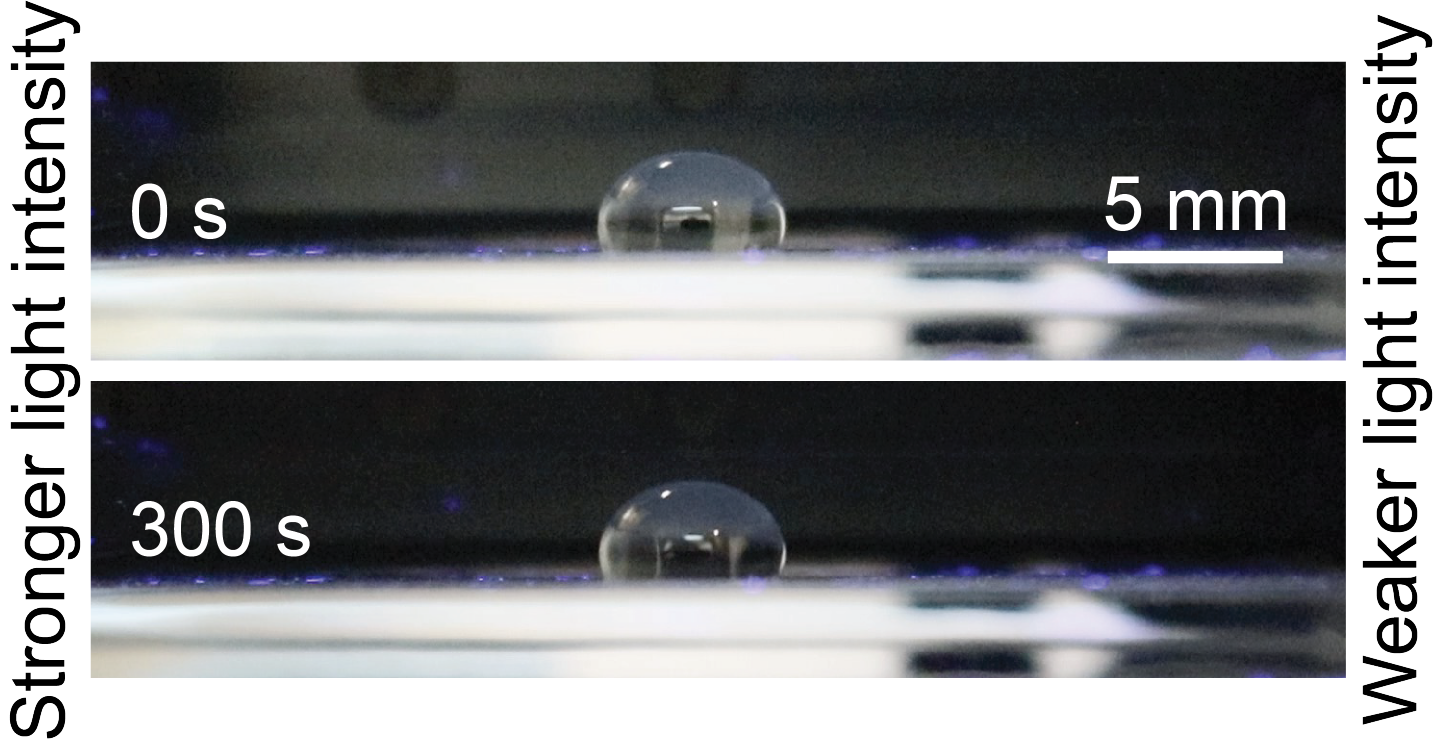}\\
Figure S15: {Optical images (side view) of a stationary pure DI water droplet illuminated by UV light with an intensity gradient after passing through an ND filter. The intensity of the light is greater on the left and weaker on the right. }
\label{Fig. S15}
\end{figure}

\end{document}